\pdfoutput=1 
\documentclass[useAMS,usenatbib,onecolumn]{mn2e}
\usepackage{psfig}  
\usepackage{graphicx} 
\usepackage{dcolumn}  
\usepackage{bm}       
\usepackage{amssymb,amsmath}  
\usepackage{color}
\usepackage{float}
\newcommand{\beq}{\begin{equation}}
\newcommand{\eeq}{\end{equation}}

\newcommand{\beqa}{\begin{eqnarray}}

\newcommand{\etal}{{\it et al.}}

\def\Kpc{\, h^{-1} \, {\rm Kpc}}
\def\Mpc{\, h^{-1} \, {\rm Mpc}}
\def\Gpc{\, h^{-1} \, {\rm Gpc}}

\def\kvecMpc{\, h \, {\rm Mpc}^{-1}}
\def\Msun{\,h^{-1}\,{\rm M_{\odot}}}
\def\ltsima{$\; \buildrel < \over \sim \;$}   
\def\gtsima{$\; \buildrel > \over \sim \;$}   
\def\simlt{\lower.5ex\hbox{\ltsima}}   
\def\simgt{\lower.5ex\hbox{\gtsima}}   
\def\etal{{et al. }}

\newcommand{\nc}{\newcommand}   

\nc{\de}{\delta}
\nc{\hn}{\hat{n}}
\nc{\bH}{\bar{H}}
\nc{\Ol}{\Om_{\Lambda}}

\nc{\ul}{\underline} \nc{\al}{\alpha} \nc{\g}{\gamma}
\nc{\Del}{\Delta} \nc{\e}{\textrm{e}} \nc{\eps}{\epsilon}
\nc{\lam}{\lambda} \nc{\Om}{\Omega} \nc{\Omm}{\Omega_m}
\nc{\Oml}{\Omega_\Lambda} \nc{\LCDM}{$\Lambda$CDM~} 
\nc{\ve}{\varepsilon} \nc{\mn}{{\mu\nu}} \nc{\vp}{\varphi}

\def\gsim{\; \raise0.3ex\hbox{$>$\kern-0.75em
\raise-1.1ex\hbox{$\sim$}}\; }

\nc{\Section}[2]{\section{#2}\label{#1}}   
\nc{\Bibitem}[1]{\bibitem{#1}}   
\nc{\Label}[1]{\label{#1}}   

\nc{\hq}{\hat{q}}
\nc{\hw}{\widehat{w}}

\def\ben{\begin{enumerate}}
\def\een{\end{enumerate}}
\def\bi{\begin{itemize}}
\def\ei{\end{itemize}}
\def\ee{\end{equation}}
\def\bea{\begin{eqnarray}}
\def\eea{\end{eqnarray}}

\nc{\M}{\rm{M}}
\nc{\Gpcc}{\rm{~ Gpc^3/h^3}}     
   
\def\etal{{et al. }}   
   
\def\ltsima{$\; \buildrel < \over \sim \;$}   
\def\gtsima{$\; \buildrel > \over \sim \;$}   
\def\simlt{\lower.5ex\hbox{\ltsima}}   
\def\simgt{\lower.5ex\hbox{\gtsima}}   
\nc{\w}{$w(\theta)$\ }   
\nc{\ie}{i.e., }    
\nc{\eg}{e.g., }

\input epsf

\begin{document}   
   
\title[The MICE Grand Challenge: Halos and Galaxies]
{The MICE Grand Challenge Lightcone Simulation II:\\
Halo and Galaxy catalogues}

\author[Crocce \etal]{
  M. Crocce, F. J. Castander, E.
  Gazta\~{n}aga, P. Fosalba \& J. Carretero \\
Institut de Ci\`encies de l'Espai, IEEC-CSIC, Campus UAB, Facultat de
Ci\`encies, Torre C5 par-2, Barcelona 08193, Spain}

\twocolumn   
\maketitle 

\begin{abstract}
This is the second in a series of three papers in which we
present an end-to-end simulation from the MICE collaboration, {\it the
MICE Grand Challenge} (MICE-GC) run. The N-body contains about 70 billion
dark-matter particles in a $(3 \Gpc)^3$ comoving volume spanning 5 orders of magnitude in dynamical range. Here we introduce the halo and galaxy catalogues built upon it, both in a wide ($5000 \,{\rm deg}^2$) and deep ($z<1.4$) lightcone and in several comoving snapshots. Halos were resolved down to few $10^{11} \Msun$. 
This allowed us to model galaxies down to absolute magnitude M$_r<-18.9$. We used a new hybrid Halo Occupation Distribution and Abundance Matching technique for galaxy assignment. The catalogue includes the Spectral Energy Distributions of all galaxies. We describe a variety of halo and galaxy clustering applications.
We discuss how mass resolution effects can bias the large scale $2$-pt clustering
amplitude of poorly resolved halos at the $\lesssim 5\%$ level, and
their $3$-pt correlation function. We find a
characteristic scale dependent bias of $\lesssim 6\%$ across the BAO feature for halos well above $M_{\star}\sim 10^{12}\Msun$ and for LRG like galaxies. For
halos well below $M_{\star}$ the scale dependence at $100\Mpc$ is
$\lesssim 2\%$. Lastly we discuss the validity of the large-scale Kaiser limit across redshift and departures from it towards nonlinear scales.
We make the current version of the lightcone halo and galaxy catalogue
({\tt MICECATv1.0}) publicly available through a dedicated web portal, \texttt{http://cosmohub.pic.es}, to help develop and exploit the new generation of astronomical surveys.
\end{abstract}   
\begin{keywords}
  (cosmology:)
  observations, large-scale structure of Universe,
 dark energy, distance scale
\end{keywords}

\section{Introduction}   

Over the past two decades our understanding of the Universe
has improved dramatically, in good part thanks to ground-breaking
observational campaigns \citep{1998AJ....116.1009R,1999ApJ...517..565P,2003ApJS..148....1B,2005MNRAS.362..505C,2004PhRvD..69j3501T}.
Although very successful this effort has opened the window to yet 
larger challenges that remain unresolved. For instance deciphering the
reason for the
late time acceleration of the Universe, what can result in totally new
forms of
energy or in the need to re-formulate Einstein's theory of gravity. There
is also a need to shed light into the nature of dark-matter and the
neutrino sector, and of a better understanding of the galaxy formation process.

The community has responded to these challenges with a multi-probe
approach consisting of several observational tests
carried on independently or combined. From cluster abundance and weak lensing
studies to large scale galaxy clustering including the baryon
acoustic oscillations and redshift space distortions
(WiggleZ\footnote{wigglez.swin.edu.au/},
BOSS\footnote{www.sdss3.org/surveys/boss.php},
CFHTLenS\footnote{www.chftlens.org},
DES\footnote{www.darkenergysurvey.org}, Euclid\footnote{www.euclid-ec.org},
DESI\footnote{desi.lbl.gov})
in addition to state-of-the-art supernovae and CMB experiments \citep{2014A&A...571A..16P,2006A&A...447...31A} and deep surveys such as GAMMA\footnote{www.gama-survey.org/} or
PAU\footnote{www.pausurvey.org/}, among many others.

\begin{table*} 
\begin{center}
\begin{tabular}{lcccccccccccccccc}
\hline \\
Run        &&&    $N_{{\rm part}}$ & \ $L_{{\rm box}}/\Mpc$  \ & \ $PMGrid$ &
$m_p/(10^{10} \Msun)$  & $l_{{\rm soft}}/\Kpc$ & $z_{{\rm i}}$   & $Max. TimeStep$ \\

           &&&  &  & &  & & &    \\

MICE-GC   &&&    $4096^3$  & $3072$    & $4096$    & $2.93$  & $50$  & $100$ & $0.02$       \\

           &&&  &  & &  & & &    \\

MICE-IR   &&&    $2048^3$  & $3072$     & $2048$   & $ 23.42$ &  $50$ & $50$   & $0.01$    \\ 

MICE-SHV   &&&    $2048^3$  & $7680$    & $2048$    & $366$   & $50$ & $150$   & $0.03$ \\  \\

\hline
\end{tabular}
\end{center}
\caption{Description of the MICE N-body simulations. $N_{{\rm part}}$
  denotes number of particles, $L_{{\rm box}}$ is the box-size, $PM
  Grid$ gives the size of the Particle-Mesh grid used for the
  large-scale forces computed with FFTs, $m_p$ gives the particle mass, $l_{soft}$ is the softening length,
and  $z_{in}$ is the initial redshift of the simulation. All
simulations had initial conditions generated using the Zeldovich
Approximation. Max. Timestep is the initial global
time-stepping used, which is of order $1\%$
of the Hubble time (i.e, $d \log a=0.01$, being $a$ the scale factor).
The number of global timesteps to complete the runs were $N_{steps}
\simgt 2000$ in all cases.
Their cosmological parameters were kept constant throughout the runs
(see text for details).} 
\label{simtab}
\end{table*}

\begin{table*} 
\begin{center}
\begin{tabular}{lcccccccccccccccc}
\hline \\
Mag Limit        &&    N. of Galaxies &  Sat. Fraction & $\langle
M^{all}_h \rangle$ & $\langle M^{cen}_h \rangle$ & $M^{cen}_{h,\rm
  min}$ & Red Cen. [\%] & Red Sat. [\%] \\ \\ \vspace{0.1cm}

$M_r < -19$   &&    $1.92\times10^8$  & $0.23$    & $1.0\times
10^{12}$    & $6.35\times10^{11}$  & $2.23\times10^{11}$ & $35$ & $77$       \\ \vspace{0.1cm}

$M_r < -20$   &&    $8.01\times10^7$  & $0.25$    & $2.38\times10^{12}$    & $1.57\times10^{12}$  & $6.04\times10^{11}$
& $46$ & $80$       \\ \vspace{0.1cm}

$M_r < -21$   &&    $1.42\times10^7$  & $0.24$    & $8.29\times10^{12}$    & $6.74\times10^{12}$  & $1.94\times10^{12}$
& $62$ & $87$       \\ \vspace{0.1cm}

$M_r < -22$   &&    $2.3\times10^5$  & $0.13$    & $5.62\times10^{13}$    & $5.5\times10^{13}$  & $7.94\times10^{12}$
& $85$ & $98$       \\ \\

\hline
\end{tabular}
\end{center}
\caption{Some basic properties of the MICE lightcone galaxy catalogue
  that we make publicly available with this series of papers. The
  catalogue subtends one octant of the full sky and reaches $z=1.4$ with no
  simulation box repetition. It is absolute magnitude limited in the
  r-band, $M_r>-18.9$ (corresponding to $M_h>2\times10^{11}\Msun$). The table
  lists the number of galaxies, satellite fraction, and mean host halo mass for
all galaxies (``central+satellites'') and ``centrals only'' above some
magnitude limits. Also listed is the minimum host halo mass for centrals and the fraction of
red centrals and red satellites (w.r.t. all centrals and all
satellites above the given $M_r$ cut).} 
\label{galtab}
\end{table*}

The task ahead is nonetheless very hard because these datasets will
have an unprecedented level of precision, and thus require ourselves to match
it with well suited analysis tools. In this regard, large and complex
simulations are becoming a 
fundamental ingredient to develop the science and to properly
interpret the results (e.g. see Fig. 2 in \cite{paperI}).

This paper is the second in a series of three in which we present 
a state-of-the-art end-to-end simulation composed of several steps,
with a strong focus in matching observational constrains and a
galaxy catalogue in the lightcone as an end-product. This was built upon a new 
N-body simulation developed by the MICE collaboration, the MICE {\it Grand
Challenge} run (MICE-GC), that includes about 70 billion dark-matter
particles in a box of about $3 \Gpc$ aside. Details of the N-body run are
given in Table \ref{simtab} and in the companion Paper I \citep{paperI}. The
N-body set up was a compromise between sampling the largest volume possible
without repetition, e.g. the one of the ongoing DES survey \citep{2005astro.ph.10346T},
while maintaining a high mass resolution, of $\sim 10^{10} \Msun$ (necessary
to reach the observed magnitude limits of current and some future
observations). 
The MICE-GC N-body run is introduced in
Paper I, with an elaborated discussion of the resulting dark-matter
clustering properties and the comparison with lower resolution runs.

Next we built halo and galaxy catalogues both in comoving and
lightcone outputs. By construction the galaxy catalogue matches
 observed luminosity functions, color distributions and clustering as a
function of luminosity and color at low-$z$
\citep{blanton03a,blanton05,2011ApJ...736...59Z}. Galaxy properties are then evolved into
the past-lightcone using evolutionary models. Some properties of the
resulting lightcone galaxy catalogue are given in Table
\ref{galtab}. Note that we also built galaxy catalogues for
the comoving outputs, which are very useful for some concrete studies.
The discussion of the
halo and galaxy catalogue construction, their properties and their
potential in terms of clustering studies are the subject of this paper
(Paper II). 

Lastly we used the
dark-matter distribution in the lightcone discussed in Paper I to
build all sky lensing potentials
and hence add lensing properties to the galaxies such as shear and
kappa values, magnified luminosities and positions, and ongoing work with intrinsic
alignments. The details of this
procedure, its validation and applications are the
subject of the companion Paper III \citep{paperIII}.

We make the first version of the MICE-GC lightcone galaxy and halo
catalogue ({\tt MICECAT v1.0}) publicly
available at the dedicated web portal {\tt http://cosmohub.pic.es}, with the hope that can be of value to
help develop the science, the design and the exploitation of new wide-area cosmological surveys. 

In this paper, besides the catalogue validation, we study
three concrete issues: (1) how the halo clustering on large-scales
depends on the mass resolution of the underlying N-body simulation for
fixed halo mass samples (2) the halo and galaxy clustering from small
scales to very large ones with a focus on scale dependent bias and
cross-correlation coefficients (3) limitations of the Kaiser limit in
Redshift Space and in particular the impact of satellite
galaxies in the multipole moments of the anisotropic clustering.

This paper is organized as follows: Section~\ref{sec:halocat} presents the MICE-GC
catalogue of friend-of-friends halos, the mass function determined from the
comoving and lightcone outputs, as well as some clustering properties.
Section~\ref{sec:mocks} describes the galaxy mocks built upon the MICE-GC
halos and their properties. In particular, we show predictions from
the galaxy mock for the clustering and
color distribution compared to observations at high-$z$ where these properties
were not matched by construction. In Sec.~\ref{sec:resolutioneffects}
we discuss mass resolution effects in 2 and 3 point halo clustering 
 statistics. In Sec.~\ref{sec:bias} we investigate the scale
dependence of bias for several halo and galaxy samples. In
Sec.~\ref{sec:rsd} we turn into redshift space and study the
applicability of the large-scale Kaiser limit in the lightcone,
and the generation of non-trivial multipole moments due to satellite
motions within halos. Finally, Sec.~\ref{sec:release} describes our public catalogue release
and Sec.~\ref{sec:conclusions} summarizes our main results and conclusions.

\section{Halo Catalogues}
\label{sec:halocat}

One of the fundamental data products of the MICE-GC simulation are
halo catalogues, which we have built out of both, comoving and
lightcone dark-matter outputs subtending the full sky. 

We identified halos using a Friend-of-Friend (FOF) algorithm with a standard linking
length of 0.2 (in units of the mean inter-particle distance) both for the
comoving outputs and the lightcone (as the mean matter density is
independent of redshift). We used a FoF code built upon the one publicly available at 
{\tt www-hpcc.astro.washington.edu}, with some concrete improvements
needed to  handle large amount of data in due time (each MICE-GC output is about
1TB of data). 

The resulting halo catalogues contain basic halo information
as well as positions and velocities of all the particles forming each halo.
This allowed us to measure also halo 3D shapes and angular momentum.
This is a key ingredient for a number of further applications, but in
particular it will permit to incorporate intrinsic alignments in
the lensing catalogues discussed in the companion Paper III.

In what follows we validate the abundance and
clustering of the MICE-GC halo catalogues. In Sec.~\ref{sec:resolutioneffects} we go one step
further and compare the clustering 
results with those of
previous runs to investigate mass resolution effects in halo bias.

\subsection{Mass function}
\label{sec:massfunction}

\begin{figure}
\begin{center}
\includegraphics[width=0.45\textwidth]{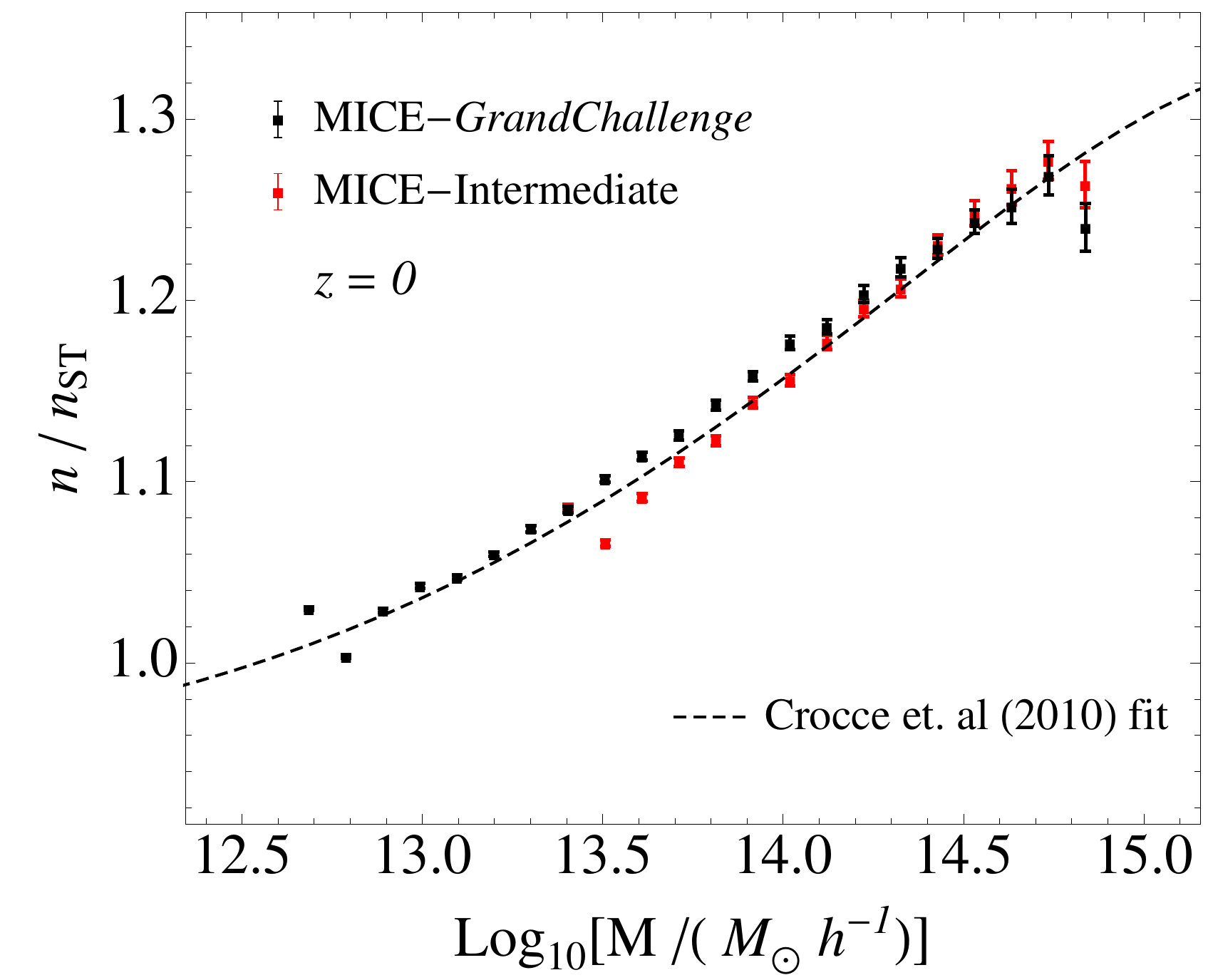} 
\caption{Mass Function in MICE at $z=0$. Black symbols show the halo
  abundance in the {\it Grand Challenge} run while red symbols correspond to the intermediate
  simulation with the same cosmology but 8 times worse mass
  resolution ($m_p = 2.9 \times 10^{10} \Msun$ vs. $m_p = 2.3 \times
  10^{11} \Msun$). The dashed line results from a fit to a series of
  MICE simulations (Crocce et al. 2010). Abundances
  are depicted relative to the Sheth and Tormen model and error-bars were estimated
  using jack-knife resampling.} 
\label{fig:MassFunctionz0}
\end{center}
\end{figure}

Let us begin by looking at the halo abundance. The halo finder in MICE-GC
yields about 172 million FoF halos with 20 or more particles in
the comoving output at $z=0$, 
and about 74 millions in each octant of the full sky lightcone up
to $z=1.4$ (these numbers rise to 350 million and 157 millions respectively if we
instead consider 10 or more particles, as we do in Sec.~\ref{sec:mocks}). Figure~\ref{fig:MassFunctionz0}
shows the measured halo mass function at $z=0$ in MICE-GC compared to both the
one measured in MICE-IR and the numerical fit to a large set of MICE
simulations from \cite{crocce10} (we depict the ratio to the well
known \cite{1999MNRAS.308..119S}  mass function  to highlight details). For each simulation, we only show a
mass-range in which halos are well sampled containing $\sim 150$ or more
particles. At the high-mass end, were halos are very well sampled in
both MICE-GC and MICE-IR, both mass functions agree very well. Towards
the regime of $ M / (\Msun)\sim 10^{13.5}$ to $10^{14}$ MICE-GC has a
slightly larger halo abundance, by $\sim 2\%$ (a trend that continues to
lower mass, not shown in Fig.~\ref{fig:MassFunctionz0}). Within $2\%$
the fit from \cite{crocce10} reproduces the shape of the MICE-GC mass
function.

We note that in defining our halo masses we have accounted for
the Warren correction for discrete halo sampling\footnote{As previously remarked this is mostly an
  empirical correction, neglecting details on other
  quantities such as halo concentration \citep{2009ApJ...692..217L}.},
unless otherwise stated (\cite{2006ApJ...646..881W,crocce10,2011ApJ...732..122B} and
ref. therein).
 This means that $M_h=n_h m_p (1-n_h^{-0.6})$, with $m_p$ being the 
particle mass and $n_h$ the corresponding number of particles in halo.
This brings the shape of both mass functions into a much better
agreement across the mass range shown in
Fig.~\ref{fig:MassFunctionz0}. 
In addition MICE-IR have been corrected for transients as
described in \cite{crocce10} (this correction is negligible for $M/(\Msun) \sim 10^{13.5}$ and
$\sim 5\%$ by $M/(\Msun) \sim 10^{15}$).
Lastly we have also accounted for the fact that the initial transfer
function in MICE-IR was EH
instead of CAMB, see Fig. 5 in Paper I (this introduces a $\lesssim 1\%$ correction,
depending on halo mass)\footnote{This was done multiplying the
  MICE-IR measurements by the mass function model prediction for the
  CAMB MICE-GC power spectrum over the one for EH used in MICE-IR.}. 

\begin{figure}
\begin{center}
\includegraphics[width=0.45\textwidth]{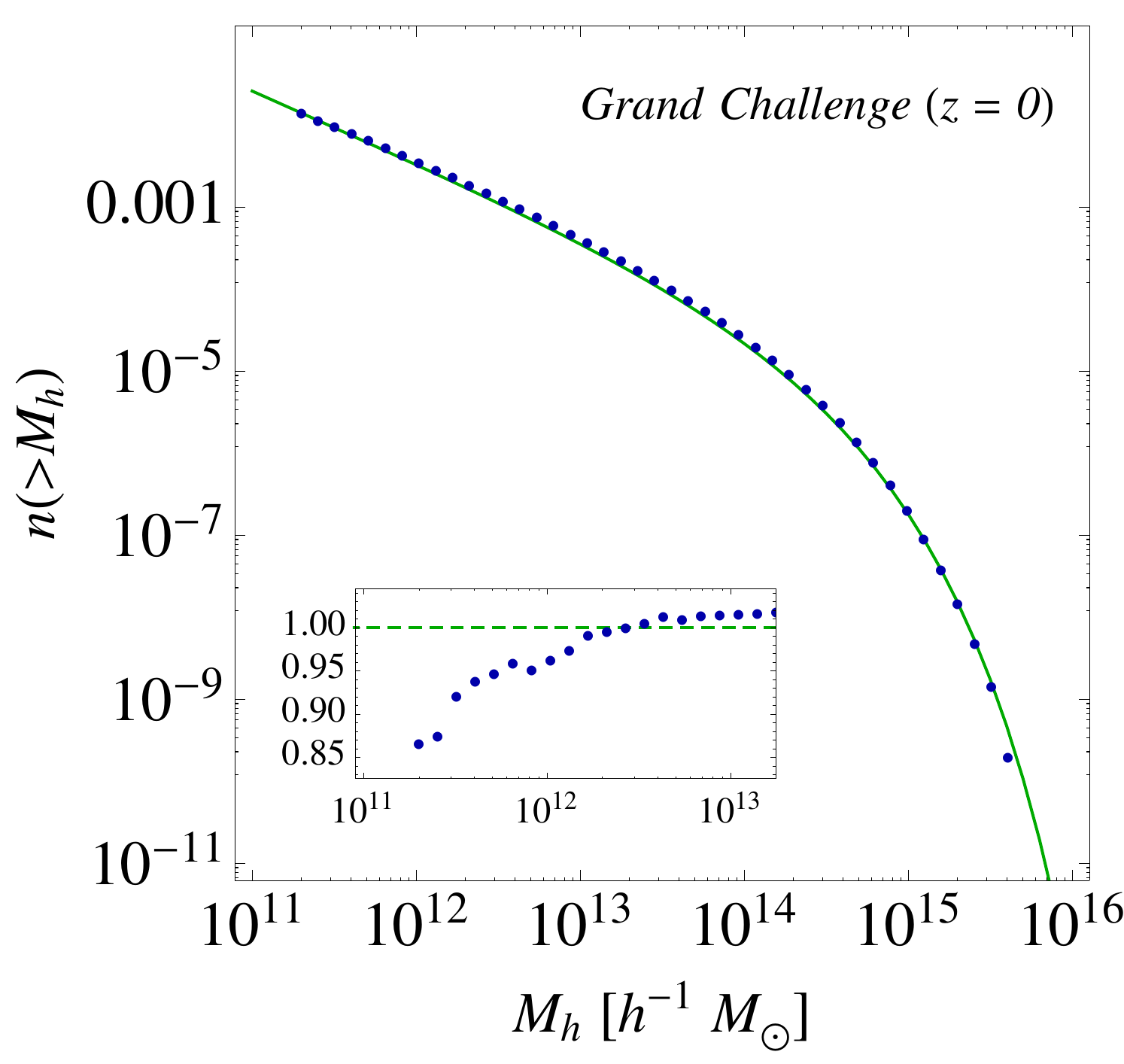} 
\caption{Cumulative mass function measured in MICE-GC down to the
  extreme regime of poorly resolved halos with 10 or more
  particles. The inset panel show the ratio to the prediction for this
quantity using the fit from Crocce et al. (2010), which is depicted by
a solid green line in the main panel.} 
\label{fig:massfunction10particles}
\end{center}
\end{figure}

Later we will argue that one possible way
  of reaching
fainter magnitudes when building galaxy mock catalogues is by using
poorly resolved structures (halos of $\gtrsim 10$ particles). It is
then interesting to investigate the abundance of these
objects. Figure~\ref{fig:massfunction10particles} shows the
cumulative mass function measured in MICE-GC down to 10 particle halos ($M_h = 2.2 \times
10^{11}\Msun$) compared to the model prediction
using the \cite{crocce10} fit \footnote{In \cite{crocce10} it is shown that their fit works
at the percentage level on this mass range.}. The cumulative abundance
is $10\%-15\%$ lower than the prediction at this limit, but it goes to
within $5\%$ for halos with $\sim$ 40 particles already. 

\begin{figure}
\begin{center}
\includegraphics[width=0.45\textwidth]{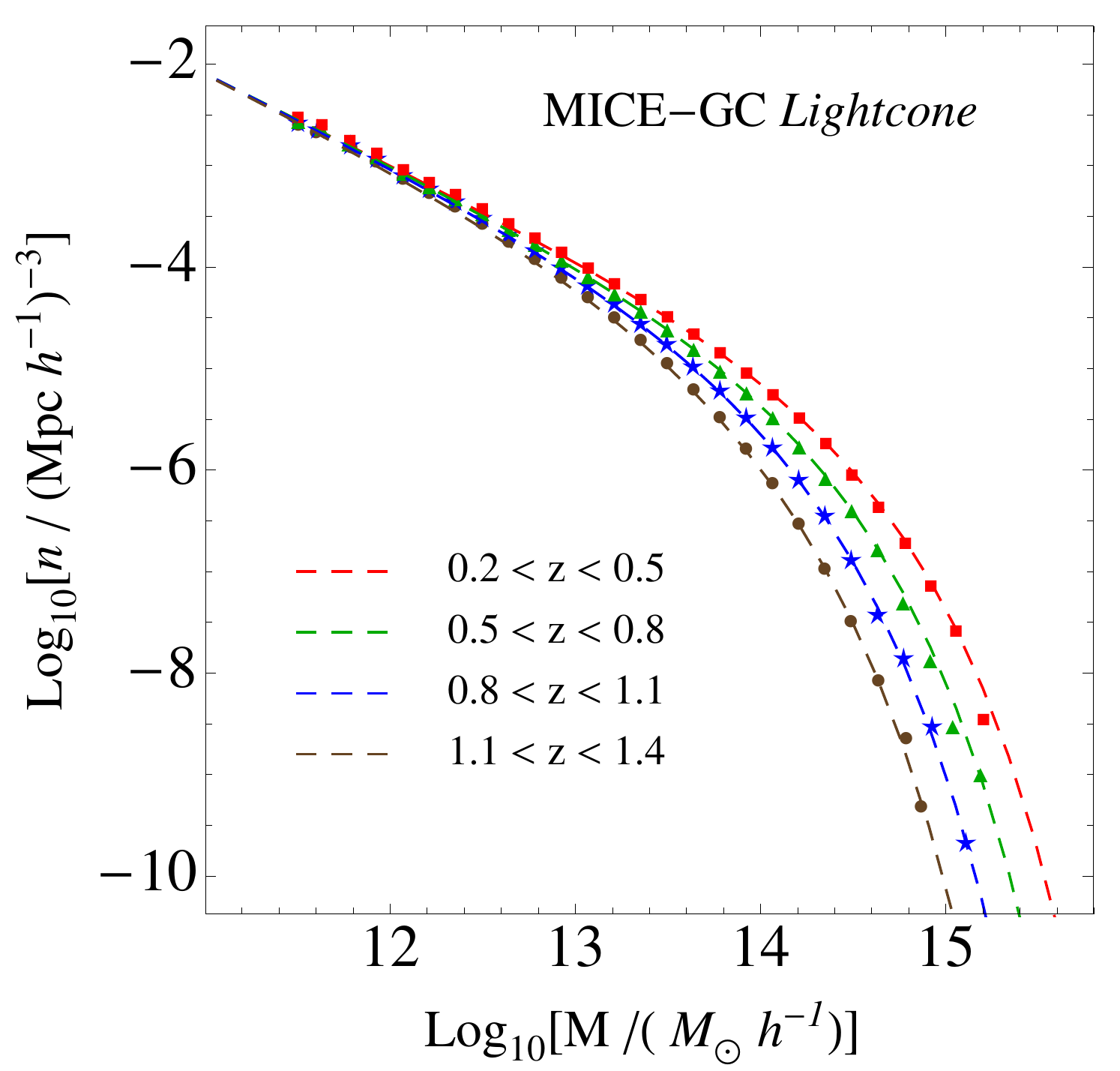} 
\caption{Halo abundance as a
  function of redshift in the MICE-GC lightcone for several consecutive redshift bins. The
  theory model shown, from Crocce et al. 2010, does not assume universality.} 
\label{fig:MassFunctionLC}
\end{center}
\end{figure}

In turn the redshift evolution of the MICE-GC halo abundance is shown in Fig.~\ref{fig:MassFunctionLC},
with the halo mass function measured in the lightcone for several consecutive
redshifts bins, as detailed in the inset labels. The evolution is in
good agreement with the fit from \cite{crocce10}, which does not
assume universality. Notice how $M_\star$ (roughly the mass beyond which the
abundance is exponentially suppressed) decreases with redshift, as
expected in hierarchical clustering (e.g. \cite{2002PhR...372....1C}).

\subsection{Halo Clustering at large scales}

\subsubsection{Power Spectrum: Clustering amplitude at $z=0$}
\label{sec:halobiaspk}

We now discuss some basic characterization and validation of the halo
clustering in the comoving snapshot at $z=0$, with a more in depth analysis postponed to Secs.~\ref{sec:bias}
and \ref{sec:rsd}. 

\begin{figure*}
\begin{center}
\includegraphics[trim = 0cm 1cm 0cm 1cm, width=1\textwidth]{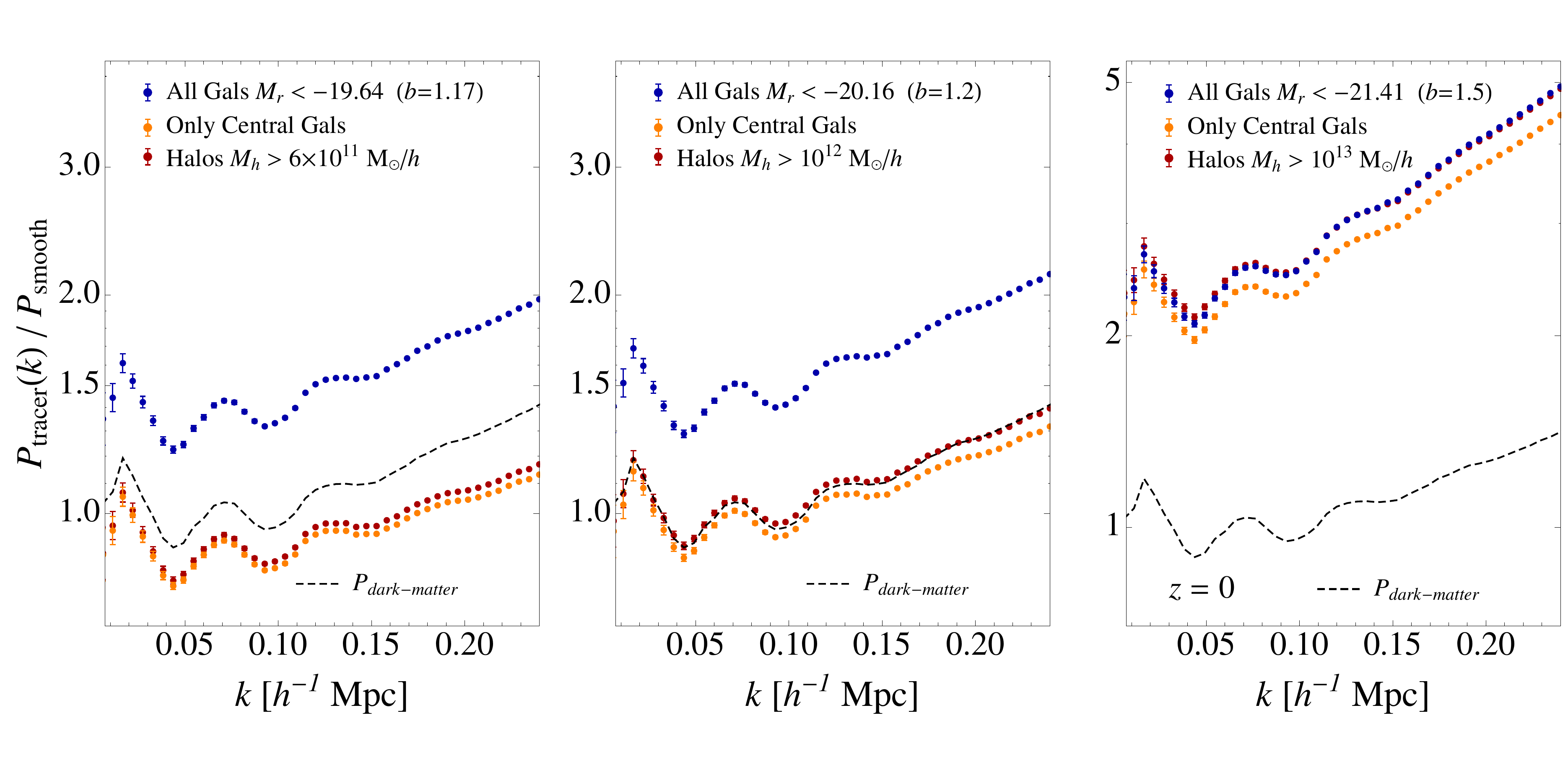}
\caption{Large-scale halo and galaxy (auto) power spectrum in the MICE-GC
  comoving output at $z=0$ (over
  a smooth broad-band power, without shot-noise correction). We display
  three self corresponding magnitude and mass threshold samples. For a given
  halo mass threshold we select the corresponding magnitude limited galaxy sample from
  the mean ``halo mass - central galaxy luminosity'' relation. We then
  consider both ``centrals only'' or ``central+satellites'' in each
  sample. The dashed black line corresponds to the clustering of dark matter.
The figure shows that the large volume and good mass-resolution in MICE-GC
allows to study large-scale clustering from anti-biased or un-biased
samples to highly biased ones, with percent level error-bars at BAO
scales. In addition notice how in general satellite galaxies
increase the clustering amplitude above the ``centrals only'' without
introducing noticeable extra scale-dependence.} 
\label{fig:halobias}
\end{center}
\end{figure*}

The combination of large volume and good mass resolution of MICE-GC allows to
study with high precision a range
of quite different halo samples in terms of clustering, from anti-biased and un-biased to
highly biased ones. The red symbols in Fig.~\ref{fig:halobias} show the $z=0$ halo-halo power
spectrum for 3 such samples in MICE-GC ($b_h=0.95, 1$ and $1.5$ in left, middle and
right panel respectively), and how precise they trace the BAO
feature. The black dashed line corresponds to the measured dark-matter clustering, which
on these scales agrees with perturbation theory predictions (RPT, \cite{RPT})
and numerical fits \citep{takahashi12,heitmann13} to $2\%$ or better, see Paper I. 
 All measured spectra in Fig.~\ref{fig:halobias} have been divided by a smooth broad-band power to highlight
narrow band features, and are {\it not} corrected by
shot-noise. Reported error bars assume the FKP approximation
\citep{1994ApJ...426...23F}: $\sigma_{P_k} =\sqrt{N_m(k) / 2} (P_k+1/\bar{n})$, where
$N_m$ is the number of Fourier modes used to measure the band-power
$P_k$ and ${\bar n}$ the tracer comoving number density.

One basic validation is the comparison of the bias measured from $P_{hh}/P_{mm}$ to the
prediction using the peak background split argument (PBS)
\citep{1986ApJ...304...15B,1989MNRAS.237.1127C}, employing the
mass function fit to MICE runs from \cite{crocce10} as input to the PBS
formulas (following Eq. (23) in \cite{2010MNRAS.402..589M}). For the samples in Fig.~\ref{fig:halobias} we find,  
\begin{eqnarray}
{\rm Halo \, Sample \ \ \ \ \ \ \ } &  {\rm b_{measured}} & {\rm bias_{PBS}} \nonumber \\
\ \ \ \ M_h/(\Msun) > 6 \times 10^{11} & 0.948  \pm 0.002 &  \ \ 0.93   \nonumber \\    
M_h/(\Msun) > 10^{12}          & 1.002  \pm 0.002 &  \ \ 0.98    \nonumber \\ 
M_h/(\Msun) > 10^{13}          & 1.546  \pm 0.009 &  \ \ 1.4 \nonumber
\end{eqnarray}\newline

\begin{figure}
\begin{center}
\includegraphics[width=0.4\textwidth]{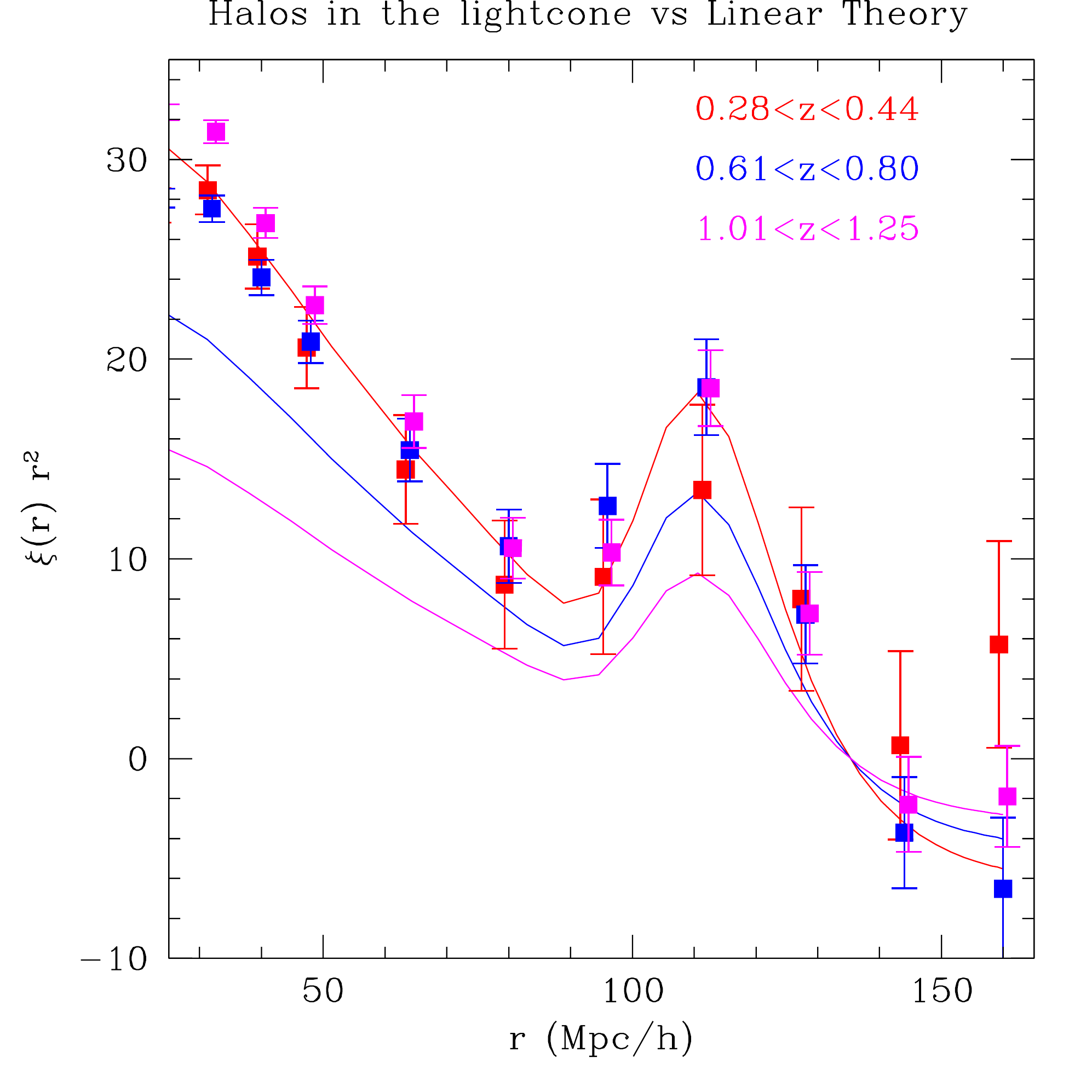}
\caption{ Symbols show the 2-point
  correlation function $\xi(r)$ (scaled by $r^2$) in FoF halos (with
  20-50 particles) for three redshift bins in
 the lightcone (in real space). Dashed lines are the corresponding DM smoothed
linear theory predictions (which resemble non-linear predictions). Notice how halo biasing for constant mass
is roughly degenerate with growth, yielding a constant clustering
amplitude.
}
\label{fig:x2h}
\end{center}
\end{figure}

The reported measured bias and errors were obtained from combining
$P_{hh}$ and $P_{mm}$ into Eqs.~(16) and (17) of \cite{2007PhRvD..75f3512S} for the first $10$ $k$-bins at the largest scales ($0.05 < k/\kvecMpc
< 0.06$). The PBS prediction agree at the $10\%$ level for the most massive
sample, in agreement with
previous studies \citep{2010MNRAS.402..589M,2010ApJ...724..878T}\footnote{Notice however that we refer to mass threshold
samples, and bias from halo-halo power spectra.}. For $b \sim 1$ samples the matching improves to $\sim 2\%$. 
To our knowledge the performance of PBS in this {\it low biasing} regime has not been explored before.

\subsubsection{Correlation Function: Halo Bias in the lightcone}

We next explore how the clustering evolves with redshift using
the lightcone halo catalogue (see also Sec.~\ref{sec:rsdgal}).  

Figure \ref{fig:x2h} shows the effect
of halo bias on the 2-point correlation at BAO scales, for a sample of 
FoF halos with masses $6\times 10^{11} < M_h/(\Msun) < 1.5\times 10^{12}$
(more than 20 particles and less than 50 particles).
We can see how the main effect of halo biasing with redshift (for
constant halo mass) is to change the amplitude of the correlations in
a way that is roughly degenerate with the linear growth in the
dark-matter (DM) distribution. The linear bias $b_1$, defined as the square root of
the  ratio of the halo correlation to the corresponding DM correlation
is about $b \simeq 0.9$ for the lower redshift and $b \simeq 1.5$ for
the largest one. These values result in halo clustering amplitudes that are quite
similar at different redshifts (i.e. $b \propto 1/D(z)$), in contrast to the corresponding DM results shown in
solid lines. 
Note that the BAO peak can be well detected despite the bias. 
We leave for Sec.~\ref{sec:bias} a more detailed study of scale dependent bias
across the BAO scales. 

Lastly
we note that the binning used in Fig.~\ref{fig:x2h} (and later in Fig.~\ref{fig:x2g}) is rather
broad to make the figure less crowded and reduce the covariance
between data-points, but we have checked that a narrower binning does not change the results.

\section{Galaxy Catalog}
\label{sec:mocks}

\subsection{Catalog construction and basic properties}
\label{sec:hod}

We have built galaxy mock catalogues from both lightcone and
comoving outputs, starting from
the corresponding halo catalogues discussed in the previous section. 
These mocks are generated to provide a tool to
design, understand and analyze cosmological surveys such as PAU, DES
and Euclid. 

We have used a new hybrid technique that combines halo occupation
distribution (HOD; e.g. \cite{jing98,scoccimarro01,berlind02}) and halo
abundance matching (HAM; e.g. \cite{vale04, tasitsiomi04, conroy06})
prescriptions. We do not intend to reproduce the details of the
catalogue generation here, but just to present some validation results
to give an idea of how the mocks compare to observations and therefore
provide a glimpse of their possible use. The galaxy assignment method is described
in full detail in \cite{carretero2014}. Further
details regarding redshift evolution of galaxy properties will be given in Castander et
al. (2014), in preparation.

We stress
however that our galaxy catalogue is not intended to reproduce one
given galaxy population, as typically needed for spectroscopic surveys 
(e.g. the CMASS sample for BOSS in \cite{2013MNRAS.428.1036M} and \cite{2014MNRAS.437.2594W}). 
In that sense our scope is larger and more
complex as we aim to reproduce the abundance and clustering across
luminosity and color space, and its evolution with redshift
(i.e. luminosity functions, color-magnitude diagrams, clustering as a
function of color and magnitude cuts, and more).

In order to generate the galaxy catalogue, we assume that halos are
populated by central and satellite galaxies. We assume that all halos
have one central galaxy and a number of satellites given by an HOD
prescription, which determines the mean number
of satellite galaxies as a function of the halo mass.
We assign luminosities to the central galaxies with a halo mass-luminosity relation computed
with HAM techniques, matching the abundances of the galaxy luminosity
function and the halo mass function. We need to introduce scatter in
this relation in order to fit the galaxy clustering at bright
luminosities. We then populate the halos with
the number of satellites given by the HOD parameters that fit the
resulting 2-point projected correlation function of galaxies to
observations. We assign luminosities to the satellites to preserve the
observed galaxy luminosity function and the dependence of clustering
with luminosity, imposing the additional constraint that the luminosity of
satellites in a given halo cannot exceed the luminosity of the
central galaxy by more than 5\% \citep{carretero2014}.


Central galaxies are placed at the halo center of  mass.
Satellite galaxies are positioned into the halos
following a triaxial NFW profile with fixed axis ratios
(e.g. \cite{jing02}). In order to fit  clustering observations as a function of
luminosity on small scales (one-halo regime) we need to concentrate
satellite galaxies more than the standard dark-matter distribution 
relation given in \cite{bullock01}. Hence we employ a concentration parameter in each axis
given by the relation in \cite{bullock01} with slight changes
depending on galaxy luminosities \citep{carretero2014}. Similar
conclusions are found in \cite{2012ApJ...749...83W}.

Central galaxies are assigned the center of mass velocity of the host
halo. In turn, we assume that satellite galaxies have in addition a virial
motion on top of the bulk halo velocity
\citep{2001MNRAS.322..901S}. We assume that the halos are virialized
and that the satellites velocities within the halo have a velocity dispersion
given by the halo mass \citep{1998ApJ...495...80B}. We assign
velocities to the satellite galaxies drawing each component of
the velocity vector from three
independent Gaussian distribution of dispersion $1/\sqrt{3}$ of the global
velocity dispersion of the halo. We discuss the impact of this
  assumption on anisotropic clustering measurements in Sec.~\ref{sec:satveldisp}.
 
\begin{figure}
\begin{center}
\includegraphics[width=0.45\textwidth]{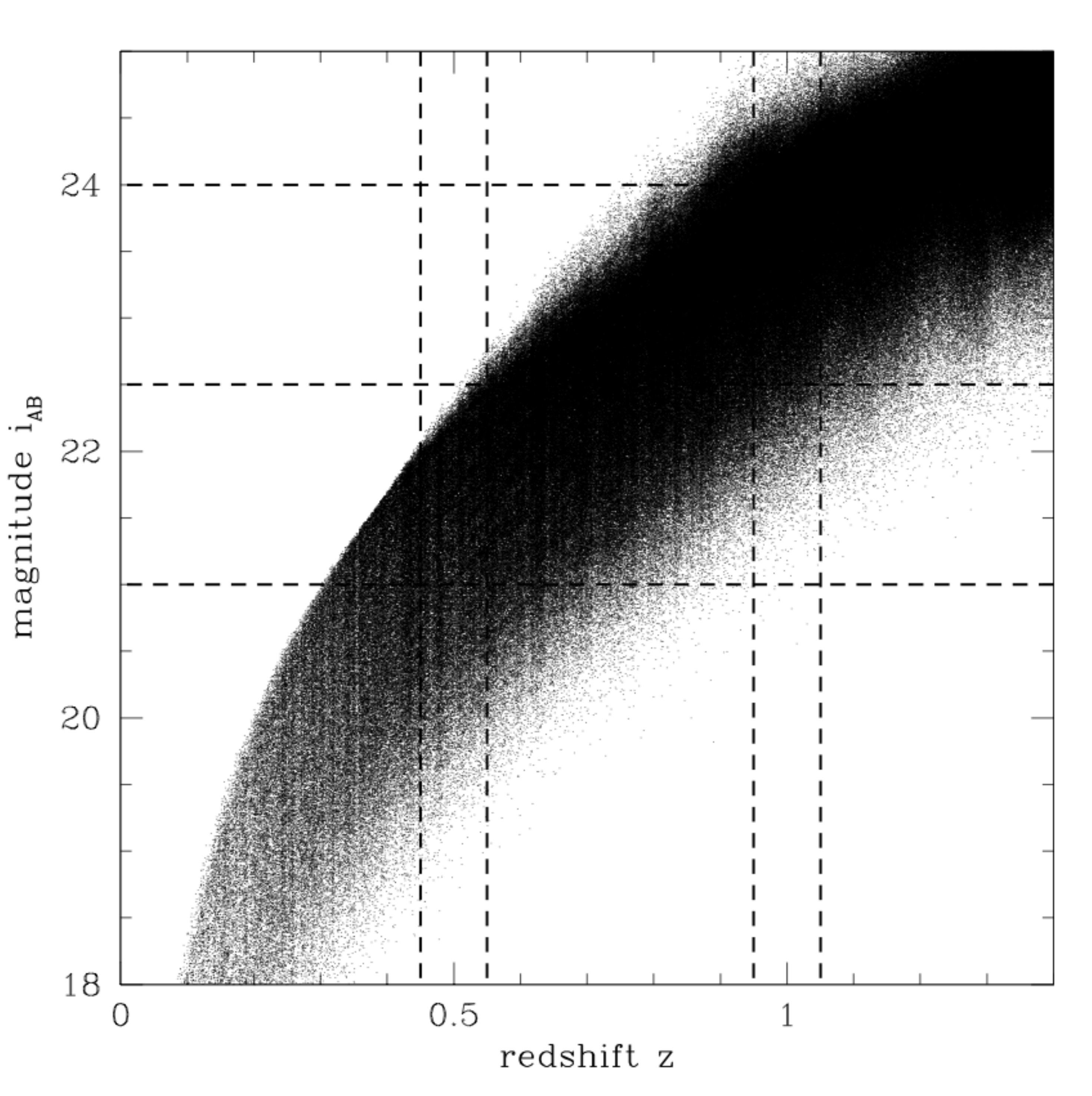} 
\caption{Scatter plot of apparent magnitudes, $i_{AB}$, versus
redshifts for a random set of galaxies in our mock catalog. 
There are missing faint galaxies at low redshift because the
sample is limited in absolute magnitude.
At $z\simeq 1$ the catalog is complete to  $i_{AB} \sim 24$,
while at $z\simeq 0.5$ is complete only to  $i_{AB} \sim 22$.}
\label{fig:plotIz}
\end{center}
\end{figure}

Lastly we assign spectral energy distributions to the galaxies with a recipe
that preserves the observed color-magnitude diagram and the
clustering as a function of color.

The method has been tuned to match observational constraints from SDSS
at low redshift where they are more stringent (detailed further below). We nevertheless apply the same
method throughout the lightcone at all redshifts with slight
modifications. In order to reproduce the observed galaxy properties at
higher redshifts, we impose evolutionary corrections to the galaxy colors
and obtain a final spectral energy distribution (SED) resampling from
the COSMOS catalogue of \cite{2009ApJ...690.1236I} galaxies with
compatible luminosity and (g-r) color at the given redshift.
Once each galaxy has a SED
assigned, we can compute any desired magnitude to compare to
observations. 

One feature of the current version of the galaxy mock catalogue
presented in this paper is that it is
absolute magnitude limited. This is inherited from the fact that is
generated from a parent halo catalogue that is mass
limited. Current cosmological imaging surveys are normally apparent
magnitude limited down to faint magnitudes. The
version of the catalogue that is now made public is complete down to
$i<24$ for redshifts $z\gtrsim0.9$. For lower redshifts the catalog is
complete only to brighter magnitudes as illustrated by Fig.~\ref{fig:plotIz}.

In order to  overcome this incompleteness at low z we have started populating
sub-resolved halos (thanks to the fact that at the resolution limit of
MICE-GC the halo bias has a weak dependence with mass) yielding
catalogues which are complete to observed magnitudes $i \sim 24$ (for
all $z < 1.4$). This additional work, more focused on modeling faint
galaxy populations at low redshift, will be presented in a follow up paper.

\begin{figure}
\begin{center}
\includegraphics[trim= 1.5cm 1.5cm 1.5cm 3cm, clip=true, width=0.50\textwidth]{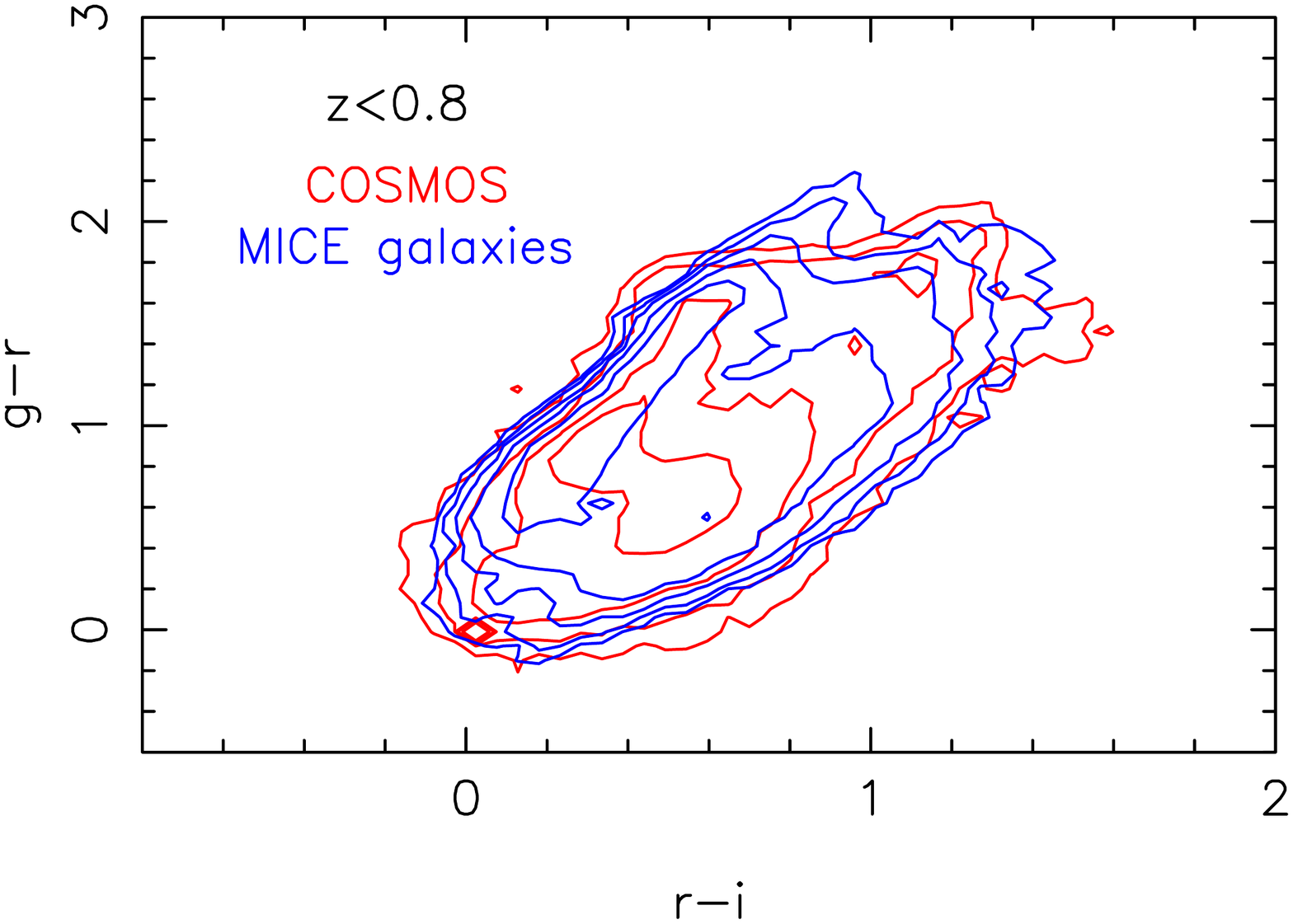} 
\includegraphics[trim= 1.5cm 1.5cm 1.5cm 3cm, clip=true, width=0.50\textwidth]{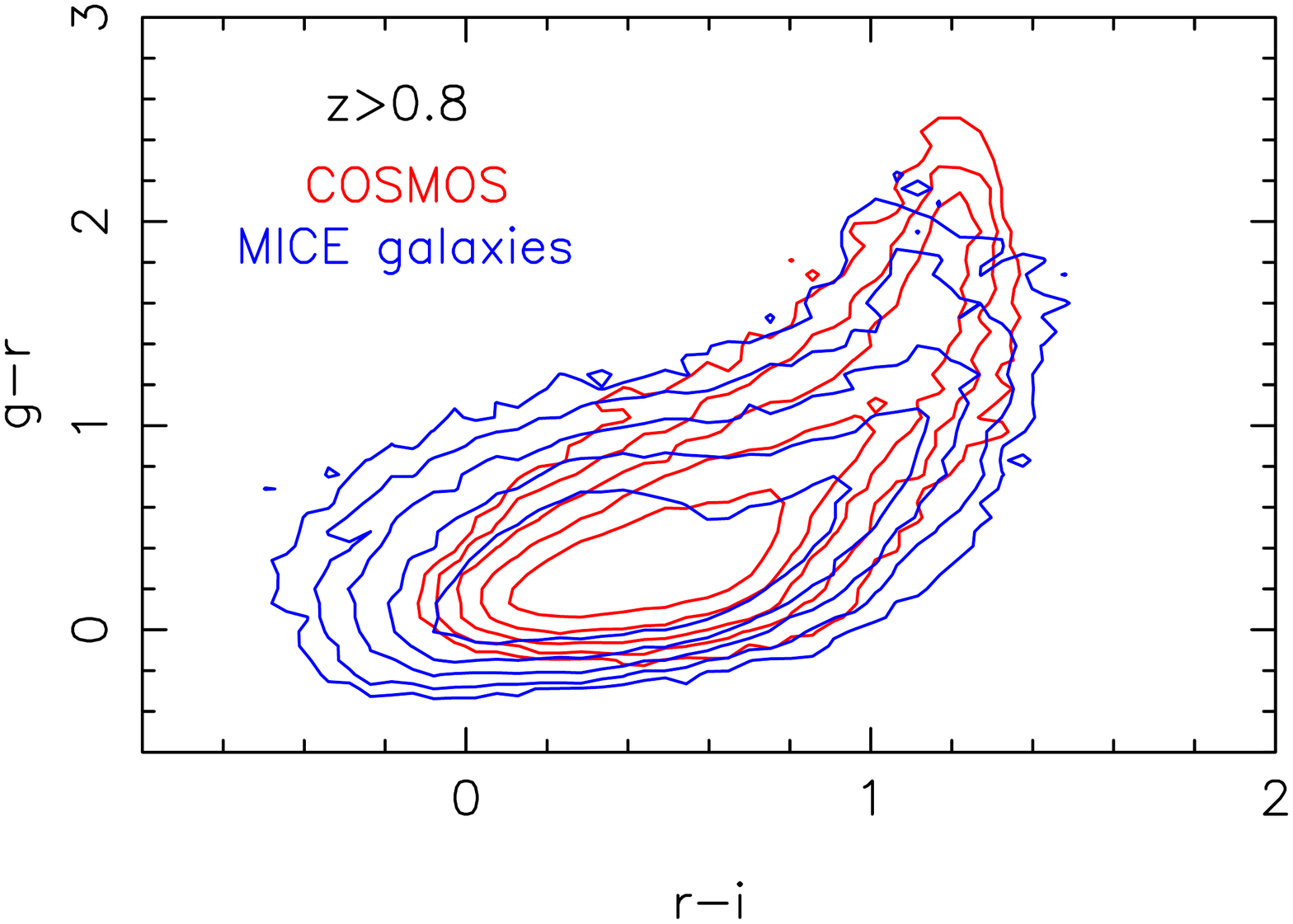} 
\caption{Contour plots for the (g-r) vs. (r-i) color distribution in the COSMOS sample
(red) and the MICE galaxy mock (blue), for galaxies at z$<$0.8 (top panel)
and between 0.8$<$z$<$1.4 (bottom panel). Notice that these distributions
are not matched by construction in our mock but rather are the result of
stellar evolution models used to populate halos in the past
lightcone.}
\label{fig:gr-ri}
\end{center}
\end{figure}

By construction our catalogue reproduces the local observational constraints
provided by the Sloan Digital Sky Survey (SDSS, \cite{york00}). We
reproduce the local luminosity function \citep{blanton03a} and color
distributions \citep{blanton03b,blanton05}. The method has also been adjusted to match the
clustering as a function of luminosity and color following
observational constraints from the SDSS \citep{2011ApJ...736...59Z}.  
Hence, we now
present a comparison of the photometric and clustering properties of
the galaxy catalogue to observations at higher redshift where the
method has not been tuned. 

As an example of such comparison,
Fig.~\ref{fig:gr-ri} shows the contour plots of the (g-r) vs. (r-i)
color distribution of the COSMOS catalogue of \cite{2009ApJ...690.1236I}
and our mock catalogue. The COSMOS data has been cut in absolute
magnitude, $M_V<-19.0$, and redshift, $z<1.4$, to expand the same
ranges as the MICE catalogue. On the top panel we present the galaxies
at $z<0.8$, where the overall color-color distribution of our mock
is a reasonable fit to observations. On the bottom panel we show the
distribution of galaxies at $0.8<z<1.4$. In this case our catalogue is
also an acceptable fit to observations albeit slightly over-producing
blue galaxies.

\begin{figure}
\begin{center}
\includegraphics[trim= 1cm 1.5cm 2cm 4cm, clip=true, width=0.50\textwidth]{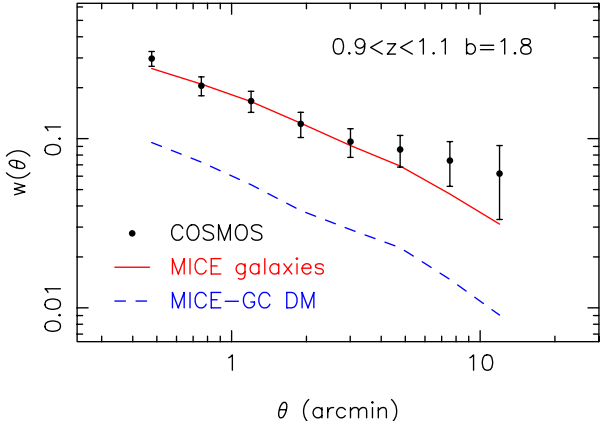} 
\caption{Angular two point correlation function of the COSMOS (black
  dots and error bars) and MICE-GC
  catalogues (red solid line) at redshift $z=1.0$ for a magnitude
  limited sample $17 < i_{AB} < 24$. We also plot the
  MICE dark matter angular correlation function (blue dashed line) for
  comparison. In this plot, the MICE correlation
  has been computed in an area sixty times that of the COSMOS
  catalogue using the same mask. Given the larger area of the MICE
  catalogue and to avoid overcrowding the plot, we do not show the MICE error bars.}
\label{fig:COSMOS_MICE_z10_wtheta}
\end{center}
\end{figure}

Turning to clustering properties, in
Fig.~\ref{fig:COSMOS_MICE_z10_wtheta} we compare the two point
angular correlation function of the COSMOS catalogue (black dots and
error bars) to our mock catalogue (red solid line) at redshift
$z=1.0$. In both catalogues, we have selected all galaxies in the
redshift range $0.9<z<1.1$ and galaxy magnitude $17.0<i<24$. The value
of the correlation amplitude is very similar, except at scales larger
than 5 arc-minutes, where the COSMOS amplitude is larger than the
catalogue (although at the 1-$\sigma$ level). The COSMOS field presents an overdensity at $z\sim0.9$ in
its redshift distribution and is observed to have somewhat larger
clustering amplitude than other fields \citep{skibba13}. For
comparison we also calculate the dark matter two point angular correlation
function (blue dashed line), from which one can infer the bias of the
sample to be $b\sim1.8$ at these scales. 

\begin{figure}
\begin{center}
\includegraphics[width=0.45\textwidth]{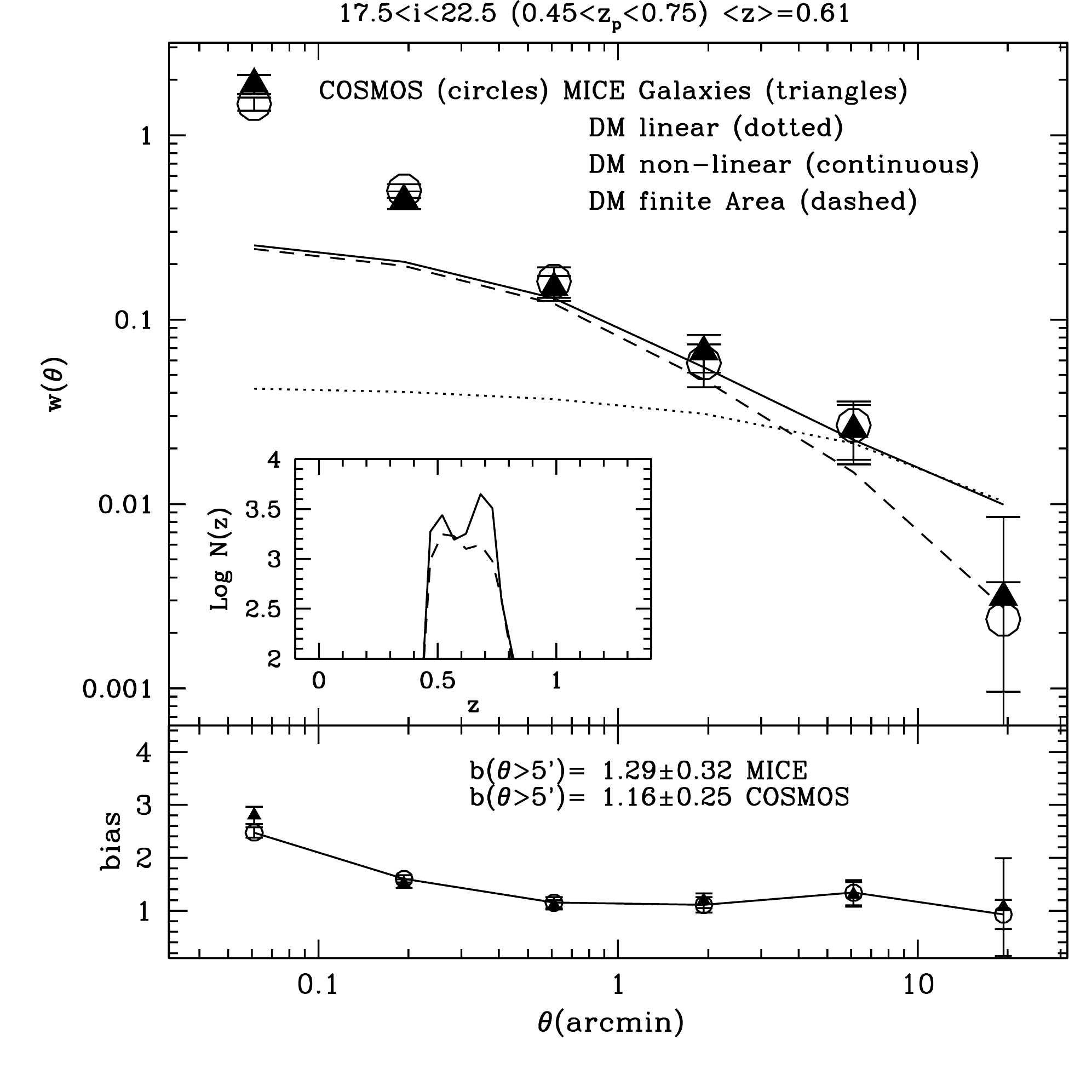}
\caption{Top Panel: Angular two point correlation function of the COSMOS
  (circles with jack-knife errorbars) at photometric redshift
$0.45<z<75$ and for a galaxy sample with $17.5 < i_{AB} < 22.5$. Triangles with  errorbars
correspond to the mean and dispersion of 50 COSMOS like MICE-GC  
catalogues. We use the same magnitude and photo-z limits 
for MICE as in COSMOS. The redshift distribution of galaxies is shown in the inset
for both COSMOS  (continuous) and MICE (dashed), with $\langle z
\rangle = 0.61$. We also plot the  DM prediction for MICE
as different lines, as indicated in the top label.
Bottom Panel: Measured galaxy bias for
COSMOS (circles) and MICE (triangles).}
\label{fig:COSMOS_MICE_z75_wtheta}
\end{center}
\end{figure}

In the top panel of Fig.~\ref{fig:COSMOS_MICE_z75_wtheta} we show the corresponding
results for a redshift slice of $0.45<z<75$. Here we
restrict both samples to brighter galaxies $17.5<i_{AB}<22.5$, so that both are
complete (see Fig.~\ref{fig:plotIz}). The COSMOS photo-z errors are also much
smaller for these brighter galaxies, so the comparison is more direct. We also
include the DM linear prediction (dotted), the non-linear DM
prediction (continuous) and the prediction including the finite area
correction (dashed line) resulting from the integral constrain in the
COSMOS area (of about $1.4 \, {\rm deg}^2$). We have simulated the COSMOS sample with 50 separate
MICE-GC regions of equal size, same magnitude limit ($17.5<i<22.5$) and
same redshifts distribution (shown in the inset).
The COSMOS data and the MICE-GC catalog agree remarkably well.
In the bottom panel we also show the effective  galaxy bias $b$ estimated
as $b(\theta)=\sqrt{w_{GAL}(\theta)/w_{DM}(\theta)}$, the ratio of the galaxy
to non-linear DM prediction. The results are quite consistent between
MICE-GC and COSMOS, despite the fact that MICE galaxy catalog was not built to
match clustering by construction at these redshifts.

\begin{figure}
\begin{center}
\includegraphics[trim = 0cm 2cm 0cm 2cm, width=0.5\textwidth]{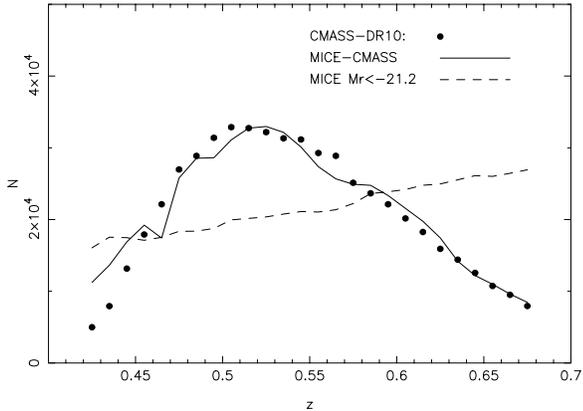}
\caption{A MICE-GC luminous red galaxy (LRG) sample. Solid black line
  (labeled MICE-CMASS) is the redshift distribution  for an
  LRG sample in the MICE-GC catalogue following the
  color and magnitude selection criteria as the SDSS-III BOSS CMASS
  sample in Anderson et al. (2012). Filled dots is the actual
  distribution from CMASS-DR10. After the color selection the redshift distribution of
  MICE-CMASS matches very well CMASS-DR10. 
  Dashed line corresponds to a sample
  selected only in absolute luminosity such as to yield the same
  clustering as DR10 (see Fig.~\ref{fig:CMASS_MICE}) and the
  same total number of objects.}
\label{fig:CMASS_nz}
\end{center}
\end{figure}

As a further validation we have compared the clustering in our
catalogue to the one in the CMASS sample of the SDSS-III BOSS survey,
on BAO scales. Note that this test
is quite more challenging than before as it involves both a
magnitude and a color selection of galaxies evolved in the past
light-cone. We built a
MICE-CMASS sample by applying the same selection for luminous red
galaxies described in Eqs.(1-6) in 
\cite{2012MNRAS.427.3435A}. In the current version of the MICE-GC
catalogue there is no evolution of the luminosity function beyond the
one of the underlying mass function. Before doing a combined color and
luminosity cut is important to account for this evolution \footnote{We
have tested that the clustering validation against COSMOS discussed
before is not affected by applying the evolution in luminosity prior to sample
selection. Mainly because the samples are not as bright as CMASS,
neither they involve a color cut.}. Hence
we first correct the absolute magnitudes according to a functional fit
derived by abundance matching between the evolving galaxy luminosity function
and the halo mass function, which was not included in the MICECAT v1.0
release, 
$M_r^{evolved} = M_r + 0.92*(atan(1.5*z) -0.149)$,
and then impose the criteria $17.5 < i < 19.9$. Next we impose the
color selection $d_\perp > 0.46$ and $i < 20.13 + 1.6(d_\perp-0.8$
where $d_\perp = r-i - (g-r)/8.0)$. Note that these values are slightly
different than the ones used in CMASS (\cite{2012MNRAS.427.3435A}). 
This is due to our galaxies
being slightly bluer in this region of color space. 
The resulting catalogue
has about the same number of galaxies ($\sim 445000$, if normalized to
the $5000\,{\rm deg^2}$ of MICE-GC) 
and a very similar redshift distribution as
the BOSS-CMASS, starting at $z \sim 0.4$ and falling off by $z \sim
0.7$. This is shown in Fig.~\ref{fig:CMASS_nz}. 
We then apply the redshift selection $0.43 < z < 0.7$, as done in 
\cite{2012MNRAS.427.3435A}.

\begin{figure}
\begin{center}
\includegraphics[width=0.45\textwidth]{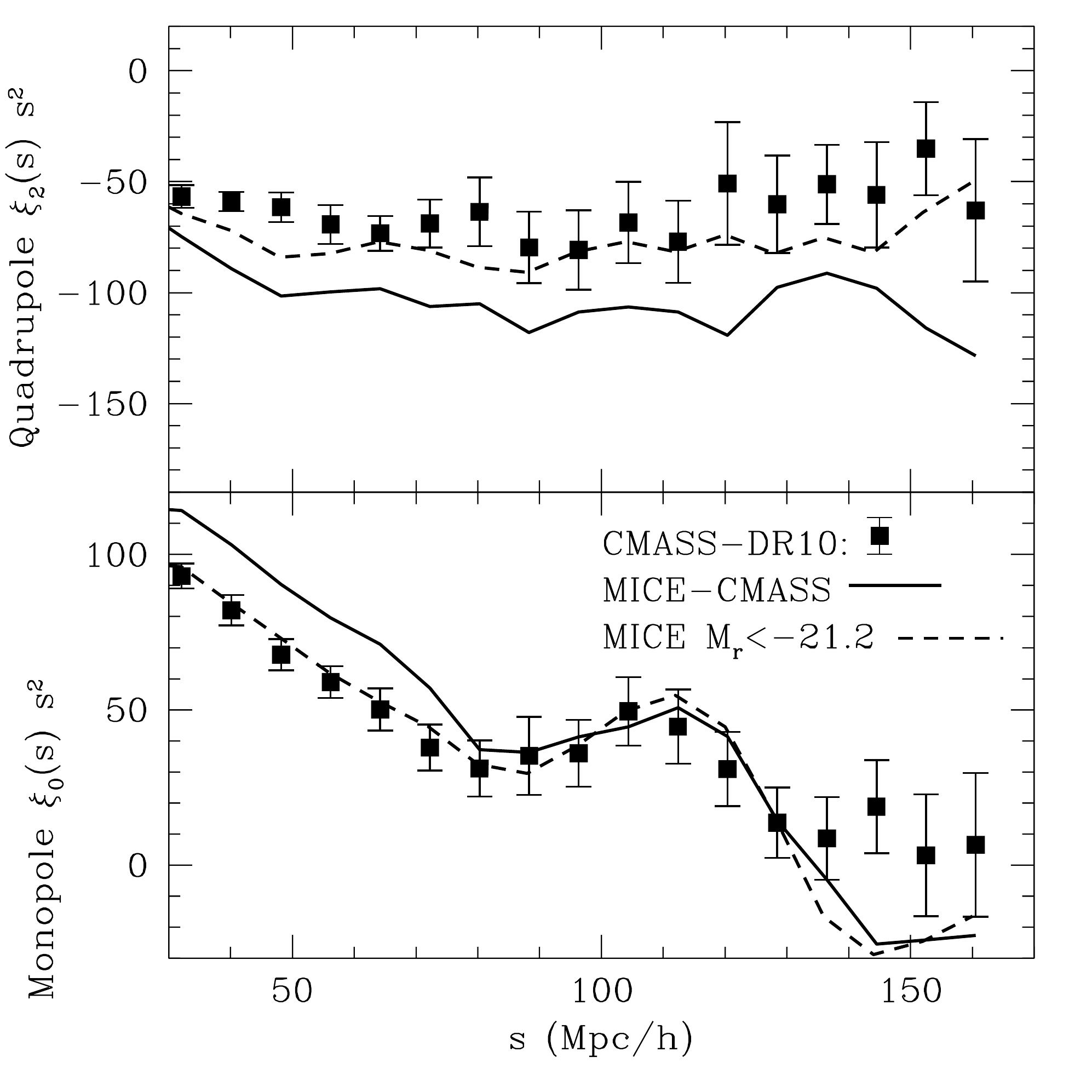}
\caption{Quadrupole and Monopole correlation functions for the
  CMASS samples built from the MICE-GC lightcone catalogue, see Fig.~\ref{fig:CMASS_nz}. MICE-CMASS is
  shown in solid black while the actual CMASS DR10 measurements with filled symbols. The
  overall shape and BAO position is well traced by MICE-CMASS, while the bias
  is $\sim 10\%$ higher (see text for details). Dashed line
  corresponds to a $M_r<-21.2$ sample in MICE-GC, that compares better
at the clustering level.}
\label{fig:CMASS_MICE}
\end{center}
\end{figure}

We next measured the monopole and quadrupole correlation functions,
focusing on large BAO scales. This is shown in Fig.~\ref{fig:CMASS_MICE} that
compares our clustering
estimate in the BOSS CMASS-DR10 sample \footnote{We have checked that
  our results are in good agreement with those in \cite{2014MNRAS.440.2692S}}
with the one over the CMASS selection of MICE galaxies.
Both the shape and BAO scale are quite well reproduced by our
catalogue. While the  relative linear bias of MICE-CMASS is 
$\sim 10\%$ higher than the one in CMASS-DR10, both in monopole and quadrupole. 
There are several factors that can contribute to this small discrepancy.
First, the MICE cosmology is different from the one in BOSS. Second,
we have fixed HOD parameters to clustering at $z \sim 0$, so we do not
expect a perfect match at higher redshift. This probably needs some HOD
evolution other than the evolution in halo properties (for example in the mass-luminosity scatter which we
apply always after the same luminosity threshold across redshift).
Third, MICE-CMASS has $27\%$ satellites which is a factor of 3 times 
larger than BOSS predictions \citep{2012MNRAS.424..136T}.  
As we fixed the 
total galaxy abundance, a lower satellite fraction requires 
including more central galaxies, which by construction have smaller halo 
mass and therefore smaller bias.  The mean halo mass in MICE-CMASS 
is $M_h=3.3\times 10^{13} \Msun$ which is a factor $2$ times higher
than that expected in 
BOSS-CMASS.
Such increase of a factor of 2 in  halo mass can yield a $\sim 10\%$ increase in halo bias, 
in agreement with our findings. An alternative sample can be
based on a simple absolute luminosity selection tuned such as to match
the CMASS clustering rather than the redshift distribution. We found
that we needed to  cut MICE-GC to $M_r < -21.2$, 
which turns out very close to the actual minimum luminosity of
BOSS-CMASS galaxies. After the luminosity selection we dilute MICE-GC to have the
same number of objects as BOSS-CMASS. The redshift distribution is
shown by a dashed line in Fig.~\ref{fig:CMASS_nz}, while the clustering in
Fig.~\ref{fig:CMASS_MICE}. The later agrees very well with CMASS-DR10
although there is no color selection imposed.

Although we have shown some concrete examples, the overall comparison
between photometric and clustering properties of our catalogue to
observations is good. This validates our approach in constructing the
galaxy mock catalogue where we have applied stellar evolutionary
corrections to the colors of the galaxies to construct the mock
catalogue in the lightcone extrapolating the other low redshift
recipes to higher redshifts. We have also discussed a simple
implementation for evolution in galaxy luminosity, particularly
relevant for narrow magnitude range selections.

\subsection{Galaxy Clustering vs. Halo Clustering}
\label{sec:galhaloclustering}

The HOD prescription described above and used to populate the MICE-GC 
simulation with galaxies is based on matching the observed luminosity 
function and the small scale galaxy clustering ($r \lesssim 30\Mpc$). In this section we
investigate what it implies for the clustering of 
galaxies on
large-scales, in particular how this compares to the halo clustering
already discussed.

Figure~\ref{fig:halobias} has three panels corresponding to the power
spectrum of
anti-biased, unbiased and highly biased halo samples discussed in Sec.~\ref{sec:halobiaspk}. In
each panel we now include the clustering of galaxies
brighter than the luminosity set by the corresponding mass-luminosity relation from
the HOD+HAM prescription. We divide the (magnitude limited) galaxy sample
into {\it centrals only} (orange symbols) and {\it central+satellites}
(blue symbols). The left-most panel shows that faint central galaxies
have almost the same clustering as their host halos. This is because
the mass-luminosity relation at this regime is one-to-one (and the
clustering is dominated by the most abundant galaxies). The addition
of satellites boost the clustering signal because faint
satellites can live in massive halos. On the other end 
bright central galaxies (right most panel) have less clustering
strength than
their corresponding halos from the mean halo mass - central
luminosity relation. This is due
to the scatter in $L_{cen}=L(M_h)$, for fixed mass. Hence, a sample
of centrals with $L > L_\star=L(M_\star)$ has galaxies residing in
halos with $M < M_\star$, what determines the (smaller) galaxy bias. 
Again adding the satellites boost the signal, in this particular case
to match that of the halos (right panel of Fig.~\ref{fig:halobias}).

\begin{figure}
\begin{center}
\includegraphics[width=0.4\textwidth]{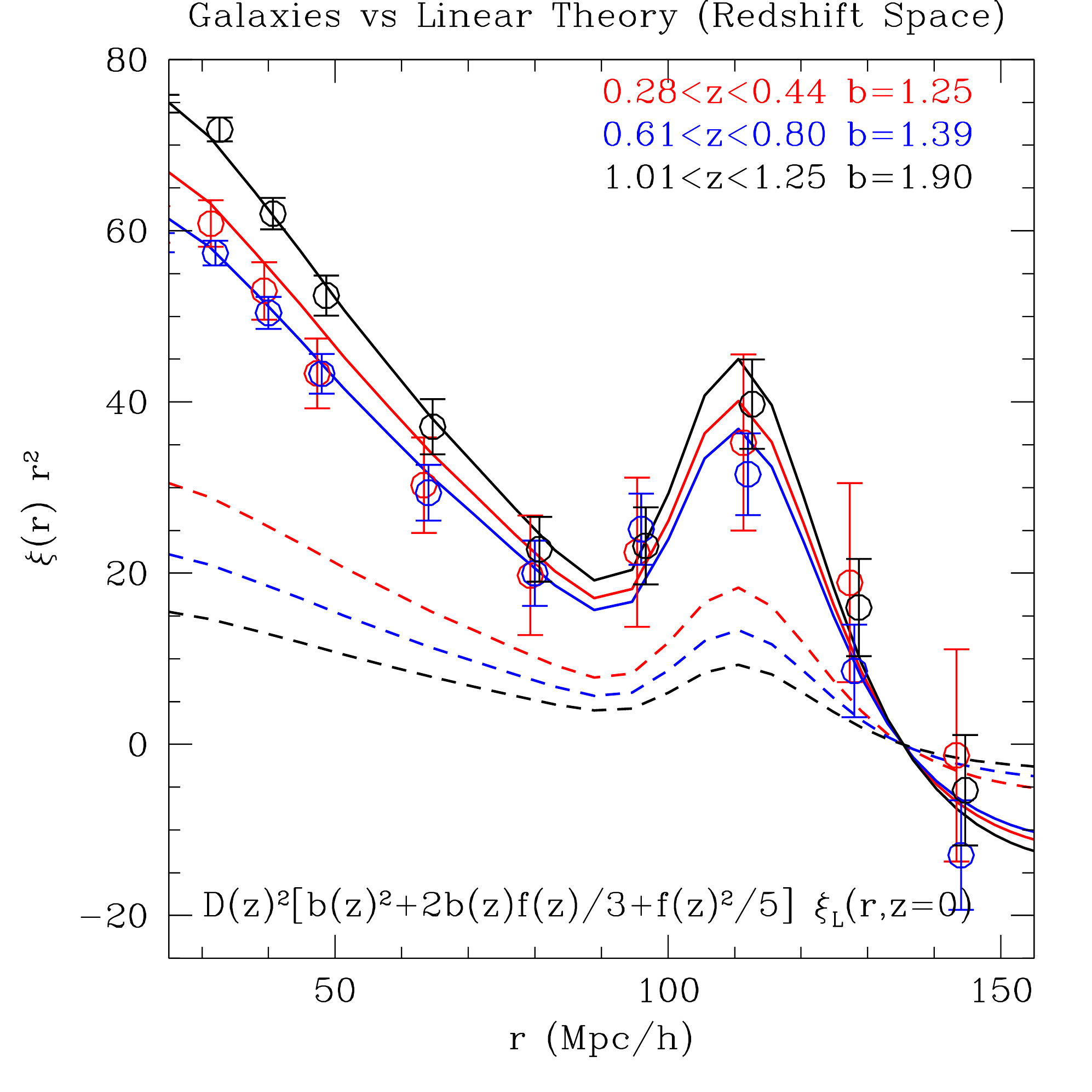}
\caption{
Galaxy 3D monopole correlation function in the lightcone (open circles) for 
three redshift bins and a magnitude limited sample $r < 24$ (to
compare with Fig.~\ref{fig:x2h}). 
  Dashed lines are the corresponding linear theory predictions
  for dark-matter in
  real-space, while solid lines include linear bias and redshift space distortions.
  The modeling, where bias has been obtained from real space
  measurements, agrees well with the galaxy monopole. In this case the
  bias evolves stronger than the growth such that the
  galaxy clustering amplitude increases with $z$, contrary to the case
of halos.}
\label{fig:x2g}
\end{center}
\end{figure}

Lastly we turn to the evolution of clustering in the lightcone (see
also Sec.~\ref{sec:rsdgal}). Figure~\ref{fig:x2g} shows the monopole 3D correlation function
measured at BAO scales in 3 redshift bins, for a
magnitude limited galaxy sample ($r<24$) extracted from one octant of the MICE
lightcone catalogue in redshift space. The dashed lines are the linear theory
predictions for the corresponding dark-matter clustering in real-space
at the given redshift. In turn the solid lines are the
linear modeling for biasing and redshift space distortions (i.e. the
Kaiser effect, \cite{kaiser84}, discussed in Sec.~\ref{sec:rsdgal} in more detailed), angle averaged 
and evaluated at the mean of each redshift bin, see Eq.~(\ref{eq:xiKai}). The bias used in the
modeling, and shown in the inset top-right labels, was obtained from
the ratio of the two point correlation of galaxies to DM in real space. As we can see, the bias
evolves quite strongly with redshift such that the clustering
amplitude is largest for higher $z$ (where dark-matter clustering is
weaker). This is because we have selected a magnitude limited
sample, hosted by halos of increasing mass as we increase the
redshift. This can be compared with Fig.~\ref{fig:x2h} that has the corresponding study with halos
of fixed mass showing a clustering amplitude that is roughly independent of
redshift, meaning $b(z) \sim D(z)^{-1}$. Overall the linear modeling
and the clustering measurements agree quite
well on large scales in the lightcone provided with the larger
statistical error bars (although more realistic from an
observational point of view) compared to a comoving output, as we
investigate in Sec.~\ref{sec:bias}. The largest differences, still at the
$1-\sigma$ level, are found at $0.28 < z < 0.44$ where sampling variance is largest and nonlinear
effects strongest. 

\section{Resolution Effects in Halo Bias}
\label{sec:resolutioneffects}

\begin{figure}
\begin{center}
\includegraphics[trim = 0cm 1.2cm 0cm 0cm, width=0.43\textwidth]{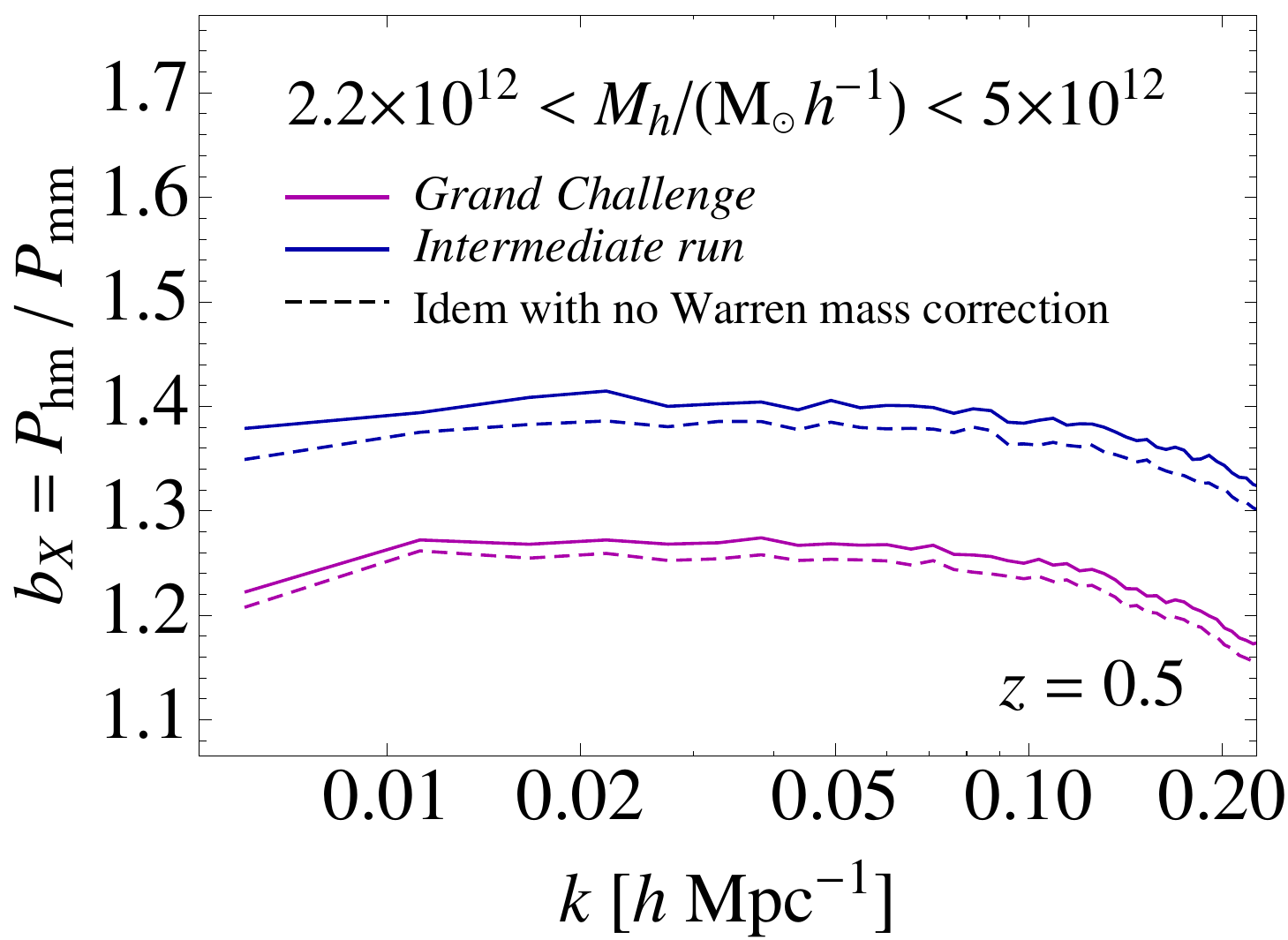} 
\includegraphics[trim = 0cm 1.2cm 0cm 0cm, width=0.43\textwidth]{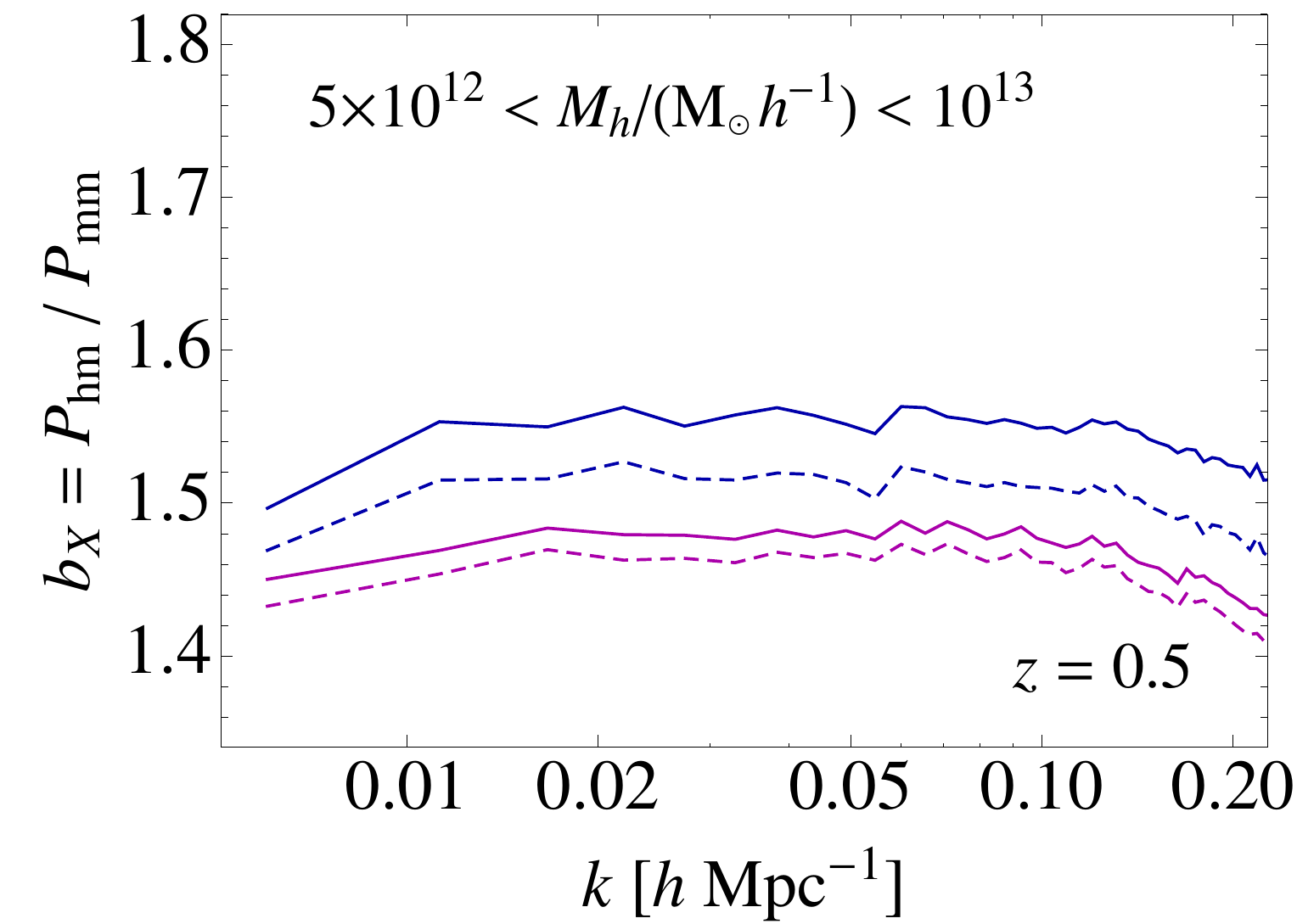} 
\includegraphics[trim = 0cm 1.2cm 0cm 0cm, width=0.43\textwidth]{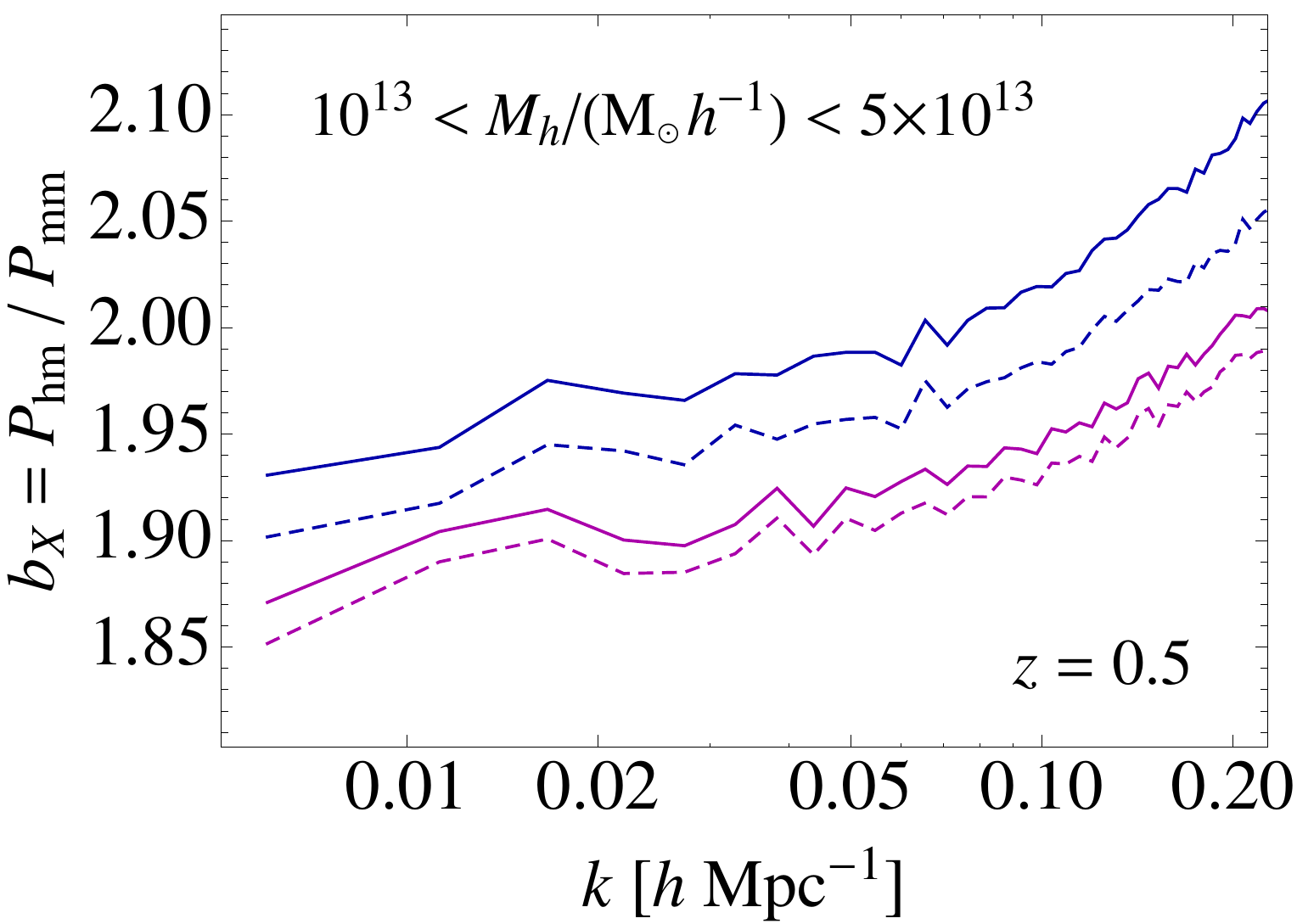} 
\includegraphics[trim = 0cm 0cm 0cm 0cm, width=0.43\textwidth]{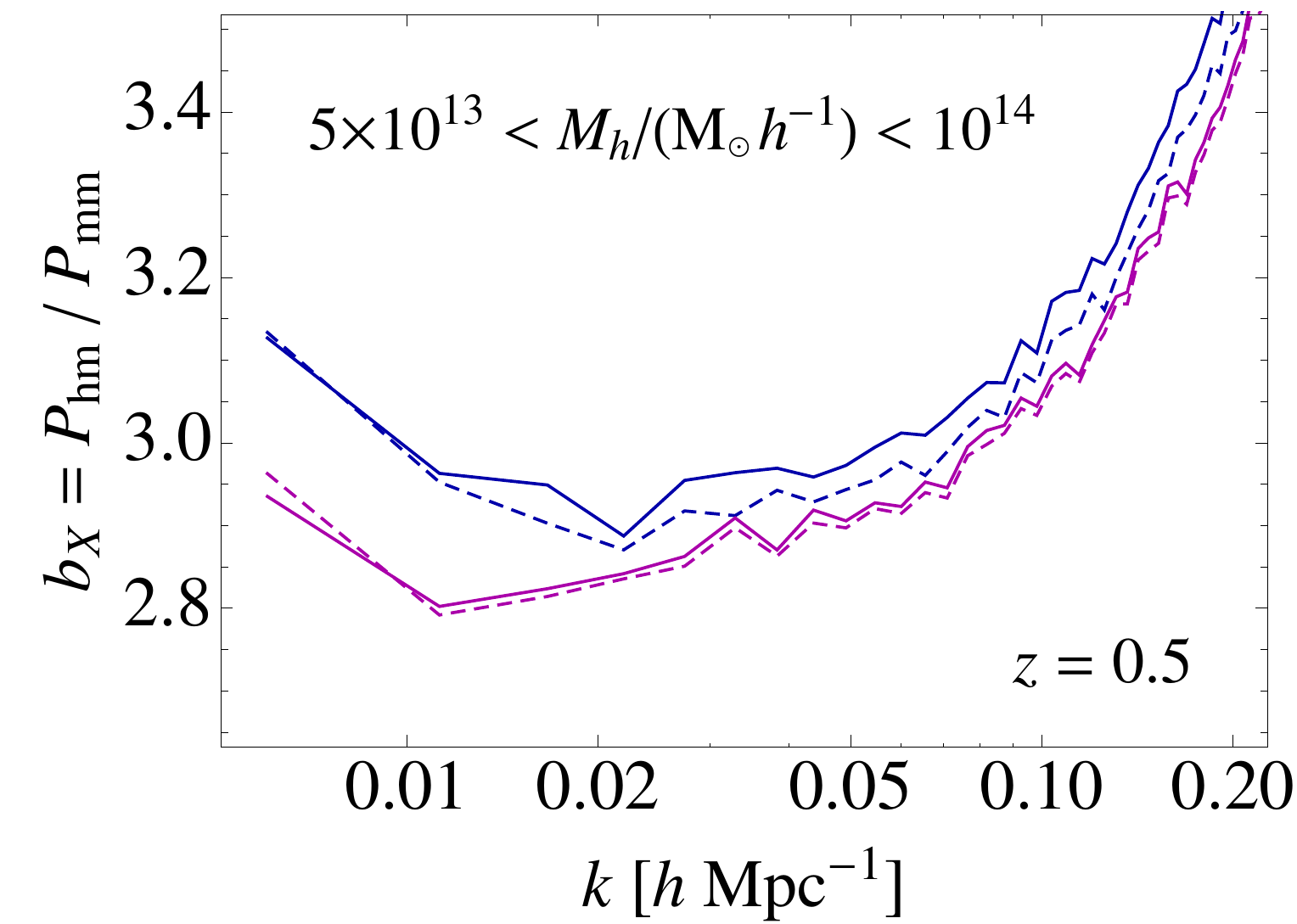} 
\caption{Mass resolution effects in large-scale halo bias. Solid
  magenta lines
  correspond to the cross-correlation bias $b_{\rm X}=P_{hm}/P_{mm}$
  measured in the MICE-GC run ($m_p=2.9 \times 10^{10}\Msun$), while
  solid blue to the one
measured in MICE-IR ($m_p=2.3 \times 10^{11}\Msun$) at
$z=0.5$. By default masses were corrected for finite halo sampling
(Warren et al. 2006). The
corresponding results for halo samples with the naive definition of mass $M_h=m_p n_p$ is depicted by the
dashed lines in each panel.} 
\label{fig:BAOh}
\end{center}
\end{figure}

In Paper I we studied how the matter distribution, in
particular the clustering, depends on the simulation mass
resolution. In this section we now extend that study to halo bias
derived from both 2-pt and 3-pt clustering.

\subsection{Results for 2-point correlations}
\label{sec:2ptmassresolution}

Let us start with the 2-pt
clustering in Fourier Space. In order to avoid
noise due to low halo densities we look at the bias through the cross-correlation
with the matter field, i.e.
\beq
b_{X} \equiv P_{hm}/P_{mm}
\eeq
where $P_{hm}$ and $P_{mm}$ are the halo-matter and matter-matter
power spectra respectively.

Figure~\ref{fig:BAOh} shows the large scale bias for different halo
samples at $z=0.5$
selected by halo mass from simulations with different particle mass
resolutions (MICE-GC and MICE-IR). 
We show $2.2 \times 10^{12}< M_h/(\Msun) < 5\times 10^{12}$, $5 \times 10^{12}< M_h/(\Msun) < 10^{13}$, $10^{13}<   M_h/(\Msun) < 5\times
10^{13}$ and $M_h/(\Msun) > 10^{14}$, from top to bottom. In each
panel MICE-GC measurements 
are depicted with solid magenta lines for samples in which halo
masses have been corrected for finite sampling prior to selection
(following the discussion in Sec ~\ref{sec:massfunction}), or with
magenta dashed lines
otherwise (here halo masses are defined simply as $m_p
n_p$). Similarly for MICE-IR we use solid blue lines for
``Warren'' corrected masses, and dashed otherwise.
In all cases the bias asymptotically approaches a scale independent value 
(although at progressively large scales with increasing halo mass, as expected) but
with some slight differences depending on the simulation and halo mass
range.

 The top panel shows an extreme case of very poorly resolved halos
(or ``groups''), formed by 10 or more particles (up to 20). At this mass scale,
MICE-GC halos have 80 to 170 particles. Even such extreme scenario
yields quite reasonable clustering, with bias miss-estimated by $\sim
10\%$. Notice that we find higher clustering amplitude in the
lower resolution simulation. We attribute this to the fact that poorly
resolved halos are found preferentially closer of big halos
and large structures in the low resolution run, that is in regions of high
density. In turn low density regions are not as well resolved as in
MICE-GC. This then biases up their clustering amplitude compared to
the same mass halos in MICE-GC.

Doing the next step down in resolution, that is, comparing the
clustering of $10$-particle halos in MICE-GC to theory predictions yield
similar or even better results, with differences at the $5\%-10\%$ level
(see Fig. 5 in \cite{carretero2014}).

As pointed in the introduction, the next generation surveys
will reach very faint magnitudes, challenging the resolution limit
of current state-of-the-art simulations. Hence different approaches
are being proposed to improve on mass resolution in approximated ways \citep{2014MNRAS.442.3256A}. This
panel intends to highlight one such approach which is simply to consider
samples of very poorly resolved halos as long as one is interested in
a halo mass scale $M_h < few \times 10^{12}\Msun$ where $b \lesssim
1$. At this scale halo bias becomes very weakly dependent on mass  (e.g. Fig 5 in
\cite{carretero2014}).
Thus for clustering measurements we can make a large error in the halo
mass and still obtain accurate results. We should stress that we make this comment with the concrete
goal of producing galaxy mock catalogues for data analysis. 
And is mainly relevant for completeness at low z because a galaxy
catalog is typically
limited in apparent magnitude therefore at high
redshifts the galaxies are quite luminous and reside only in high mass
halos. In turn error in the mass function (because of possible errors in halo mass) 
are automatically corrected by the calibration to low redshift galaxy
luminosity with the SAM\footnote{Note that SAM will correct the abundance and
  large scale clustering (two-halo term) by a suitable choice of the
  number of satellite and central galaxies at a given luminosity. This in principle modifies also the 
  one-halo term. But our satellite assignment algorithm has freedom to control the
  distribution of the satellites away from an NFW (and their
  velocities) in such a way that one can simultaneously match the
  small-scale one-halo clustering to observations \citep{carretero2014}.}.
The actual  HOD parameters will be different than in a high resolution
run, but the galaxy distribution will be quite similar. This
is why we believe that halos with small number of particles can give 
results which are similar to those in higher resolution simulations

For the next mass bin in Fig.~\ref{fig:BAOh} MICE-IR halos have $\sim
20-50$ particles (as opposed to $170$ to $350$ in MICE-GC). The
large-scale bias in MICE-IR is higher by $\sim 5\%$ if masses have
been corrected or $3\%$ otherwise (see top panel in Fig.~\ref{fig:BAOh}). The effect diminishes in the
intermediate mass range at the middle panel (with $50$ to $200$ particles in
MICE-IR halos) to $4\%$ and $2\%$ roughly. For well sampled halos
($M_h/(\Msun) > 5\times 10^{13}$, bottom panel) the bias recovered from
MICE-IR is compatible within $\sim 1\%$ to the one from MICE-GC.
For low number of particles per halo, the ``Warren'' correction seems to
introduce a mass shift too large, that translates into artificial bias
differences. A slight modification such as the one proposed in \cite{2011ApJ...732..122B}
might alleviate this issue.

In summary the bias differences found between the different
resolutions is within $3\%$ or better for well resolved halo masses, more
standard in the literature, and up to
$10\%$, for poorly resolved ones formed by as low as 10 particles. Similar conclusions were
reached studying other redshifts. Hence this kind of effect can then be of
importance for studies of accuracy in halo bias modeling, for instance
the stated accuracy of peak-background split approach is $\lesssim 10\%$
\citep{2010MNRAS.402..589M,2010ApJ...724..878T,2011MNRAS.415..383M},
not far from the effects purely dependent on simulation parameters discussed above.

\subsection{Results for 3-points correlations}

A further quantity of interest on top of the linear large-scale bias
discussed above is the second order bias $b_2$ that naturally appears 
at the leading order in higher order correlations \citep{1993ApJ...413..447F}.

On large enough scales, where the fluctuations in the density field are smoothed so that the
matter density contrast is of order unity or smaller, one can assume a
general non-linear (but local and deterministic) relation 
between the density contrast in the distribution of halos $\delta_h$
and dark matter $\delta_m$ that can be expanded in a Taylor series
\beq
\label{deltataylor}
\delta_h=\sum_{k=0}^\infty \frac{b_k}{k!}\delta^k_m=b_0+b_1\delta_m+\frac{b_2}{2}\delta^2_m+\cdots,
\eeq
where the $k=0$ term comes from the requirement that \mbox{$<\delta_h>=0$.} Within this local
bias model, at scales where $\xi(r) \equiv \langle \delta_m^2(r) \rangle < 1 $,  we can write the biased (halo or galaxy) two and three point functions
to the leading order in $\xi$ \citep{1993ApJ...413..447F,1994ApJ...425..392F}
\bea
\xi^h(r) &\simeq& ~b_1^2 ~\xi(r) \nonumber\\
\zeta^h(r_{12},r_{23},r_{13}) &\simeq& b_1^3 \zeta
(r_{12},r_{23},r_{13}) +  \nonumber \\
&& + \, b_1^2 b_2 \left(\xi(r_{12})
\xi(r_{13} \right) + {\rm cyc})
\eea
where $\zeta$ is the matter 3-pt function, which is ${\mathcal
  O}(\xi^2)$ for Gaussian initial conditions. From the above
we obtain the reduced 3-point function $Q_3$ 
\citep{1977ApJ...217..385G}
defined as:
\beq
Q_3 \equiv \frac{\zeta(r_{12},r_{23},r_{13})}{
\xi(r_{12})\xi(r_{23})+\xi(r_{12})\xi(r_{13})+\xi(r_{23})\xi(r_{13})} 
\label{eq:reduced3PCF}
\eeq
such that 
\beq
Q_3^h \simeq {1\over{b_1}} ~\left(Q_3+c_2 \right)
\label{Q3G}
\eeq
where $c_2 \equiv b_2/b_1$, and the $\simeq$ sign indicates that this
is the leading order contribution in the expansion given by Eq.~(\ref{deltataylor}) above.
In the local bias model we can use $Q_3$ as measured in the DM field
to fit $Q_3^h$ from halos, and obtain an estimate of $b_1$ and $c_2$,
that could be used to break the full degeneracy of $b_1$ and growth
in $\xi(r)$.

\begin{figure}
\begin{center}
\includegraphics[width=0.43\textwidth]{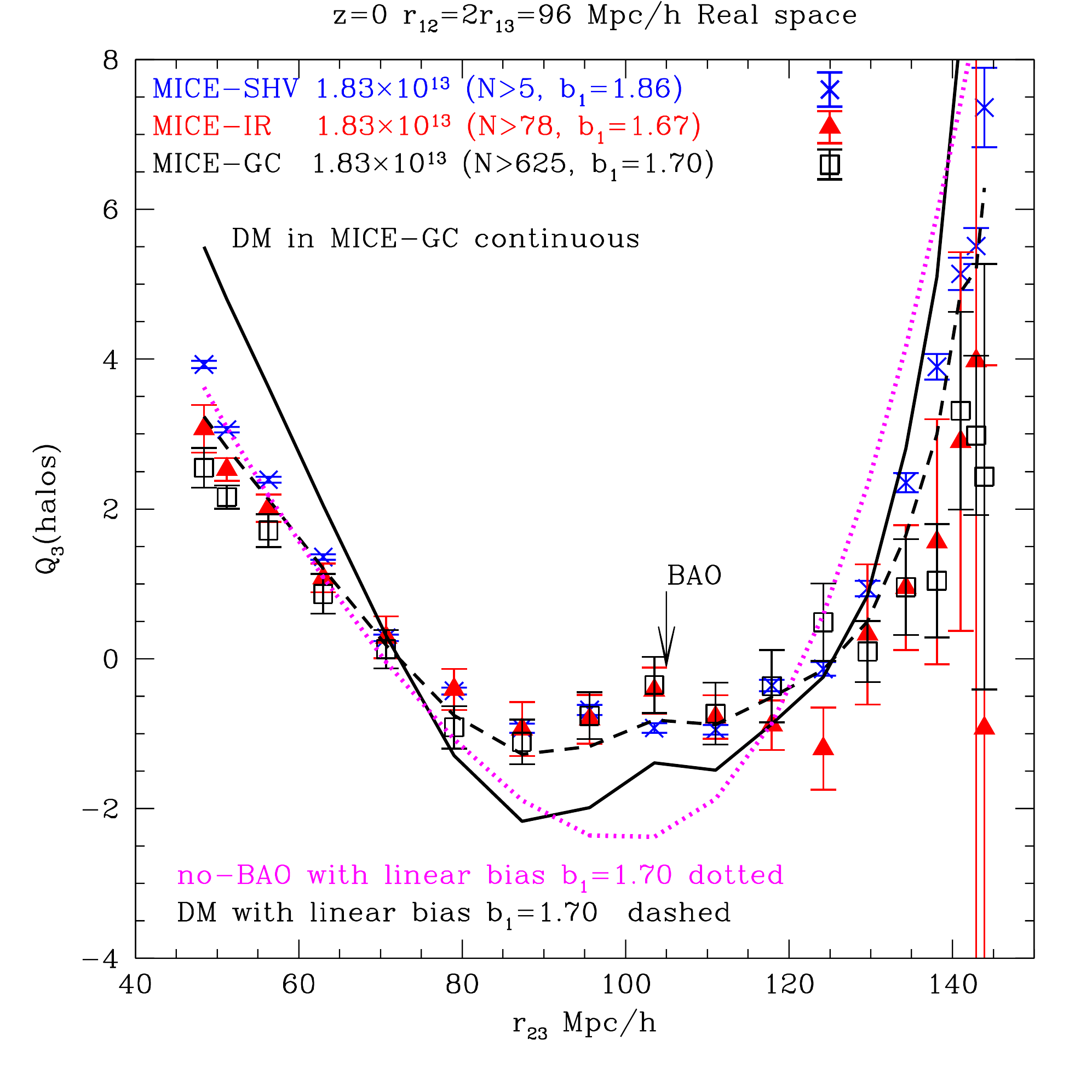}
\caption{Reduced 3-point function in Eq.~(\ref{eq:reduced3PCF}), for halos above
 $1.83\times 10^{13} \Msun$ measured in  simulations with  
different particle mass resolutions (as labeled).
We include $Q_3$ measured in the dark-matter and $Q_3/b_1$
corresponding to a local bias model (with $b_1$ estimated from 2-pt functions as in Sec.~\ref{sec:bias}).
The dotted line corresponds to the no-wiggle EH power spectrum. The imprint of the BAO
feature in $Q_3$ is clearly significant at $r_{12}\approx 110\Mpc$.}
\label{fig:q3hr96}
\end{center}
\end{figure}

Figure~\ref{fig:q3hr96} shows a comparison of $Q^h_3$
in halos of mass $M_h > 1.83\times10^{13} \Msun$ (without Warren
correction) from simulations with
different mass resolution, as detailed in Table \ref{simtab}: 
MICE-GC in black squares, MICE-IR with red triangles and MICE-SHV with
blue crosses.
The simulation with intermediate resolution has 8 times
less particles per halo than the one with higher resolution, while the
lowest resolution has 125 fewer particles. Notice that in the later case we are
using as few as $5$ particles per halo as threshold and yet $Q_3$ has a very reasonable shape.
In addition to the measured $Q^h_3$ we include $Q_3$ for the dark-matter
in MICE-GC (solid black) and a linearly biased version $Q_3/b_1$
(dashed black), see Eq.~(\ref{Q3G}). The dotted magenta is a similar
estimate but assuming a theory $Q_3$ derived from a power spectrum
with no wiggles. The figure focuses on large scales
($r_{12}=2r_{13}=96\Mpc$) and the BAO peak is clearly detected at $r_{23}\approx110\Mpc$.

\begin{figure}
\begin{center}
\includegraphics[width=0.43\textwidth]{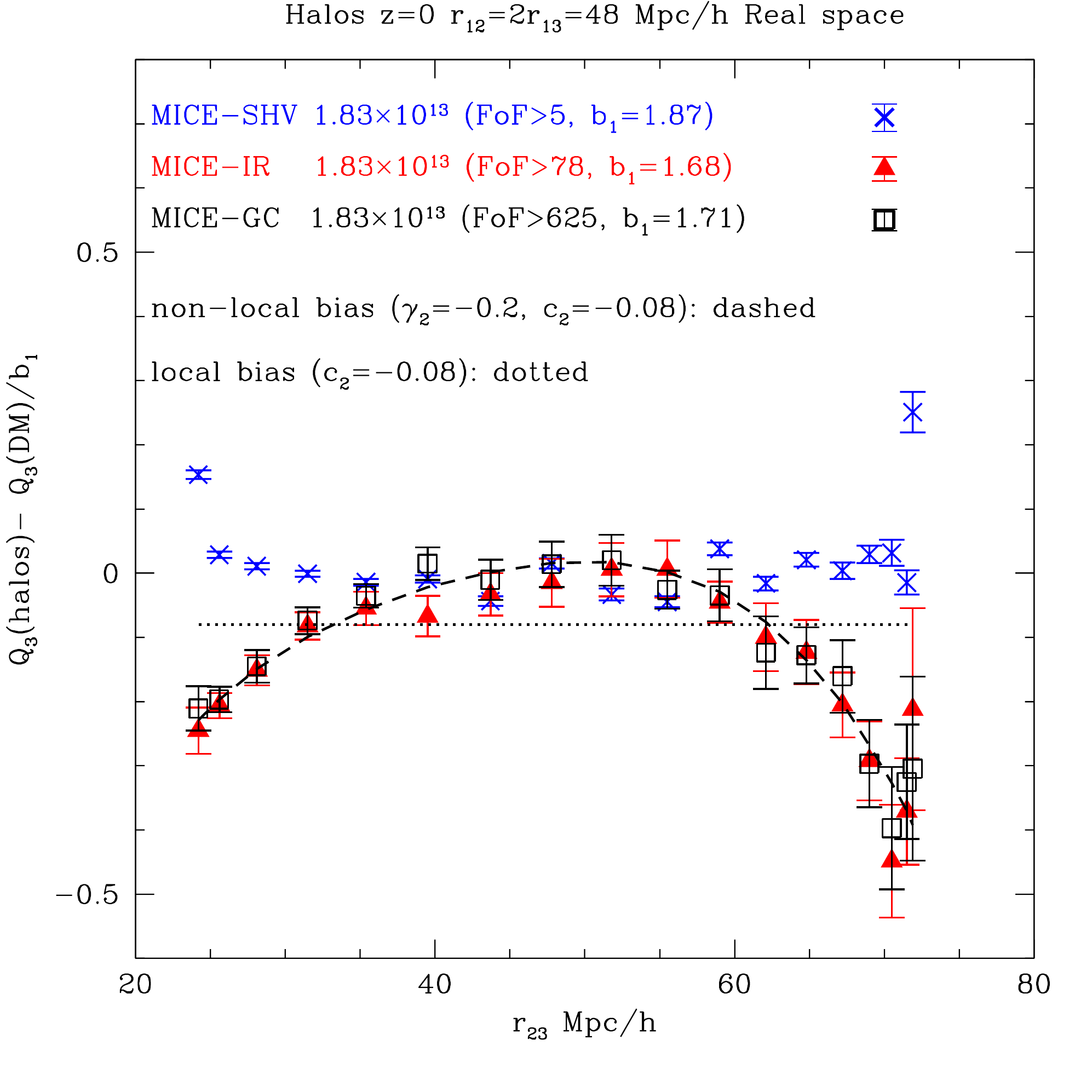}
\caption{Amplitude of the reduced 3-point function relative to the
  dark matter one measured in each run. Note how MICE-GC and MICE-IR
  agree with each other and with the non-local model. In turn, the
  MICE-SHV yield differences of few percent and seems to follow the
  local model (dotted line)}
\label{fig:q3hr48}
\end{center}
\end{figure}

The differences between MICE-GC and MICE-IR are marginal, with derived
linear bias values that agree at the percent level as found in
Sec.~\ref{sec:2ptmassresolution} for well resolved halos. In turn MICE-SHV yield larger differences showing that such a low
resolution is inappropriate for percent level accuracy studies. One
subtlety is that even at the level of dark-matter there are some
differences among these simulations, as discussed in Paper I. 
Figure~\ref{fig:q3hr48} shows the effect of resolution on nonlinear
bias by plotting $Q^h_3 - Q^{dm}_3 / b_1$ where both
$Q_3^h$ and $Q^{dm}_3 \equiv Q_3$
are measured in each given run. Thus we subtract the linear bias
which also has some resolution coming effects coming from the DM.
Moreover we focus on smaller scales,
$r_{12}=2r_{13}=48\Mpc$. 
In Fig.~\ref{fig:q3hr48} the local bias model corresponds to an
horizontal line set by the non-linear bias $c_2$ while the non-local model of
\cite{2012PhRvD..85h3509C} is given by the dashed black
line\footnote{Note that our best fit non-local coefficient $\gamma_2$
  is in perfect agreement with the one derived from $b_1=1.7$ following
  the relation found in \cite{2012PhRvD..85h3509C}: $\gamma_2 \simeq
  -2(b_1-1)/7 = -0.2$}.  
For isosceles triangles ($r_{23}=48\Mpc$) all three simulations
agree, but for collapsed and elongated shapes MICE-SHV exceeds considerably
the other two runs. In fact, while MICE-GC and MICE-IR track well the
non local model clearly deviating from a horizontal line, MICE-SHV seems to 
be consistent with it for $c_2 \sim 0$ (in addition to
$\gamma_2=0$). Further work regarding higher order halo bias in MICE-GC
computed with different methods can be found in \cite{2015MNRAS.447.1724H}.

\begin{figure*}
\begin{center}
\includegraphics[trim= 0.8cm 1.2cm 0cm 0cm, clip=true, width=0.8\textwidth]{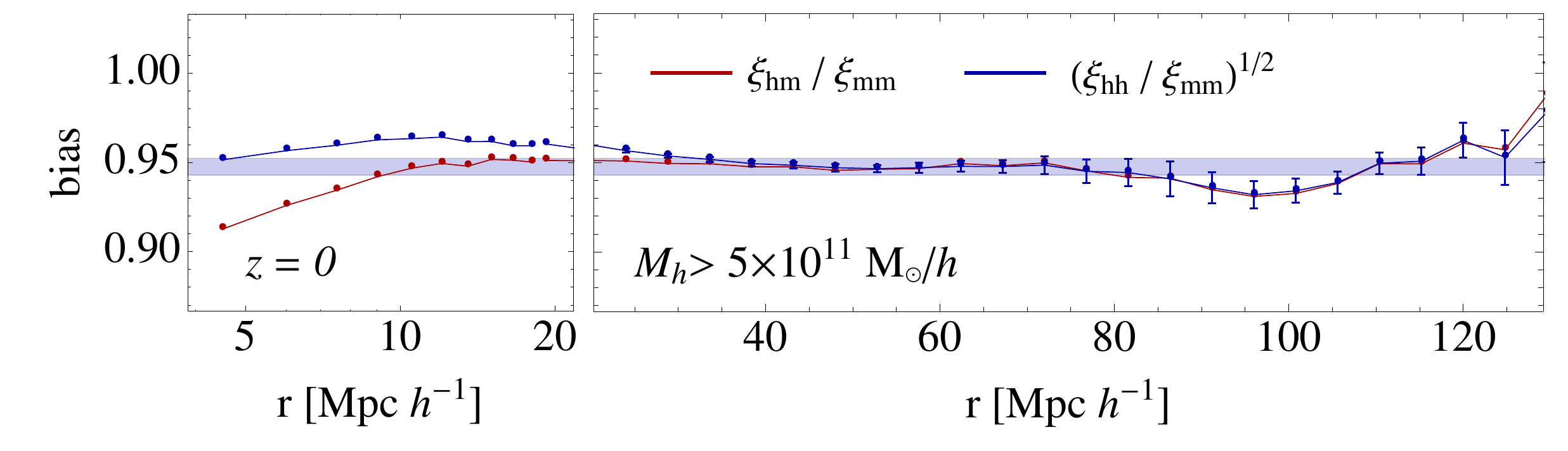} \\
\includegraphics[trim= 0.8cm 1.3cm 0cm 0cm, clip=true,width=0.8\textwidth]{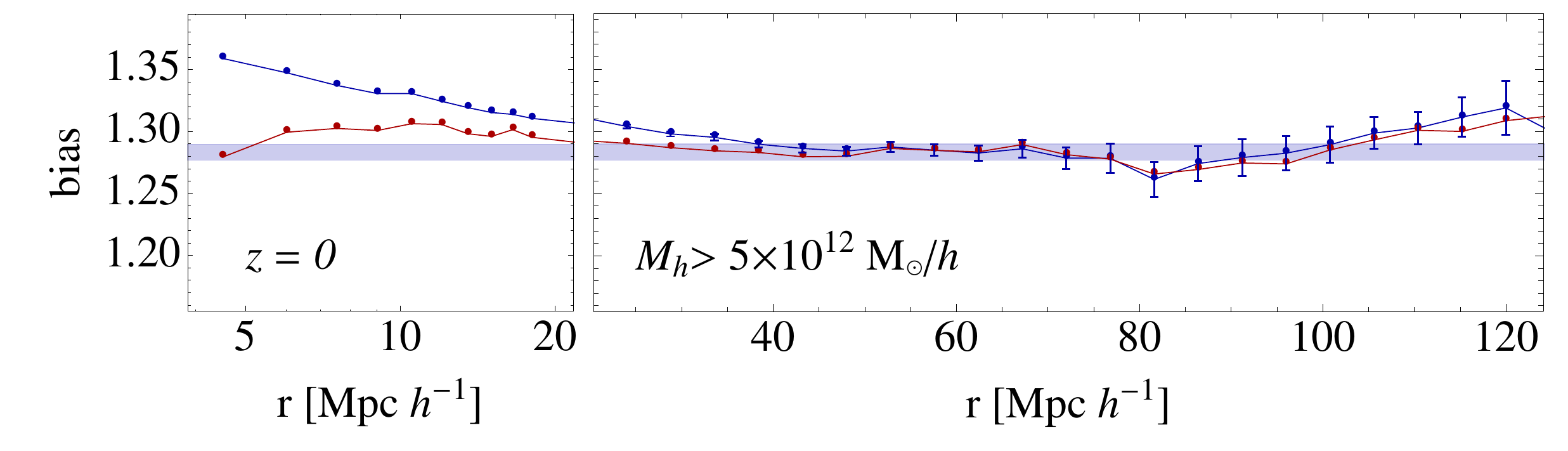} \\
\includegraphics[trim= 0.8cm 0.3cm 0cm 0cm, clip=true, width=0.8\textwidth]{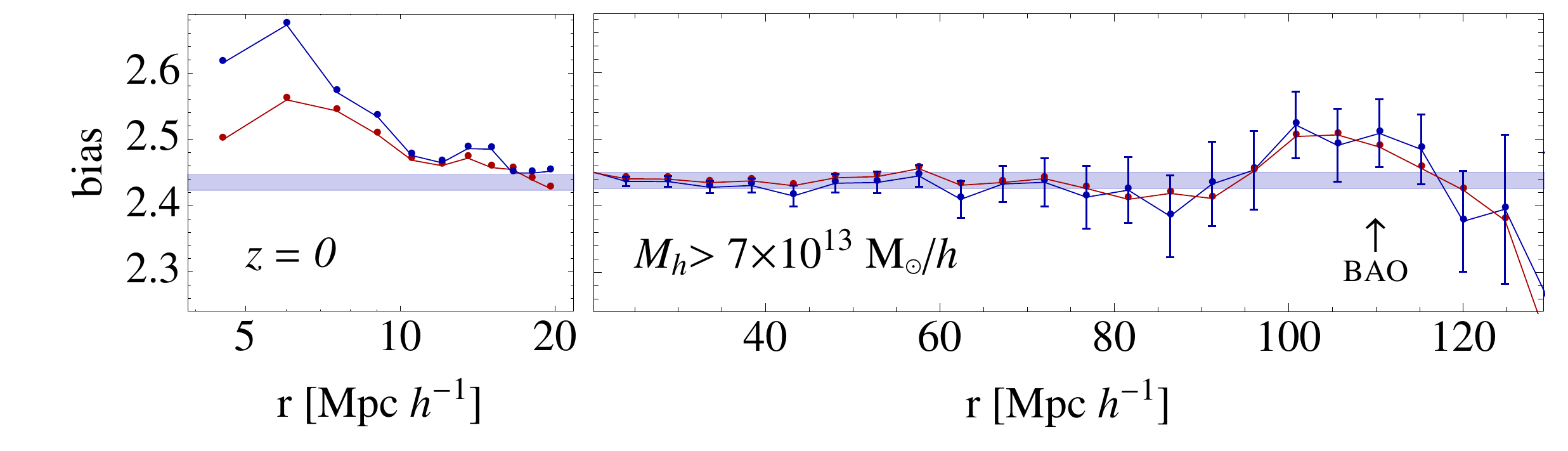} 
\caption{Halo bias in MICE-GC for 3 different mass threshold samples
  (as labeled). The shaded regions indicates a $1\%$ around the
mean value measured at $r > 30 \Mpc$. For an $M_\star$-like sample the bias from halo-matter correlations
(red symbols) is very close to scale-independent from small to large
scales. At small scales ($r \lesssim 20\Mpc$) there is a clear trend of the clustering
with increasing halo mass (with $b_{hm}\equiv \xi_{hm}/\xi_{mm}$ being
smaller, similar and larger than
the large-scale value). Across the BAO region there is slight clustering
decrement ($\sim2\%$) at $100\Mpc$ for the least massive halos and a clustering excess 
of $\sim 4-5\%$ at the BAO peak position (marked with a vertical
arrow) for the most massive ones. For clarity error bars are only displayed for bias
  derived from $\xi_{hh}$.} 
\label{fig:BAObiashalos}
\end{center}
\end{figure*}

\section{Halo and Galaxy Bias from small to large (BAO) scales}
\label{sec:bias}

As recalled in the introduction, one of the most interesting aspects
of the MICE-GC run and its derived products is the combination of
large volume and good mass resolution. In this section we profit from
these by looking in detail at the scale dependence of halo and galaxy
bias in configuration space, from small scales relevant for full-shape
fitting \citep{2009MNRAS.400.1643S,2013MNRAS.433.1202S,2014MNRAS.440.2692S} to large scales tracing the BAO
feature (\cite{2012MNRAS.427.3435A} and references therein). 

In Fig.~\ref{fig:BAObiashalos} we show the halo bias from two point correlation
functions for three ``mass threshold'' halo samples, $M_h /
(\Msun)> 5\times 10^{11}, 5\times 10^{12}$ and $7\times 10^{13}$. 
These samples were selected to have low, mid and high
bias (top to bottom panels respectively). On the one hand
Fig.~\ref{fig:BAObiashalos} focuses on the comparison of 
the bias from the halo cross-correlation signal with dark-matter
($b=\xi_{hm}/\xi_{mm}$, red line) versus the one
from halo auto-correlation ($b=(\xi_{hh}/\xi_{mm})^{1/2}$, blue line). On the other hand the panels split the bias measurement
into small scales (shown with logarithmic binning) and large-scales (shown with
linear binning). This is useful to determine on what scales the
cross-correlation coefficient $r_{cc} \equiv \xi_{hm} /
\sqrt{\xi_{mm}\xi_{hh}}$ departs from unity, a point that is typically linked
to the brake-down of a
local and deterministic biasing \citep{1998ApJ...500L..79T,1999ApJ...520...24D}. 
In each panel the filled region shows $1\%$ of
the mean linear bias defined as the error weighted average over scales
$s \ge 30 \Mpc$. Defined in this way we find $b_1=0.95, 1.28$ and $2.43$ (top to bottom).
Both, Figs. \ref{fig:BAObiashalos} and \ref{fig:BAObiasgalaxies}, correspond to comoving catalogues at $z=0$.
Error bars in those figures were obtained using jack-knife resampling with
${\rm njk}=64$ regions (measuring the bias in each region and the mean and
variance weighted by ${\rm njk}-1$ afterwards).

Overall the halo bias is remarkably close to scale independent within
few percent from large scales down to $15-20 \, \Mpc$, with the
cross correlation coefficient $r_{cc}$ being close to unity on this
regime. On the largest scales there are however some residual effects worth highlighting.

\begin{figure}
\begin{center}
\includegraphics[trim= 0cm 0cm 0cm 0cm, clip=true, width=0.42\textwidth]{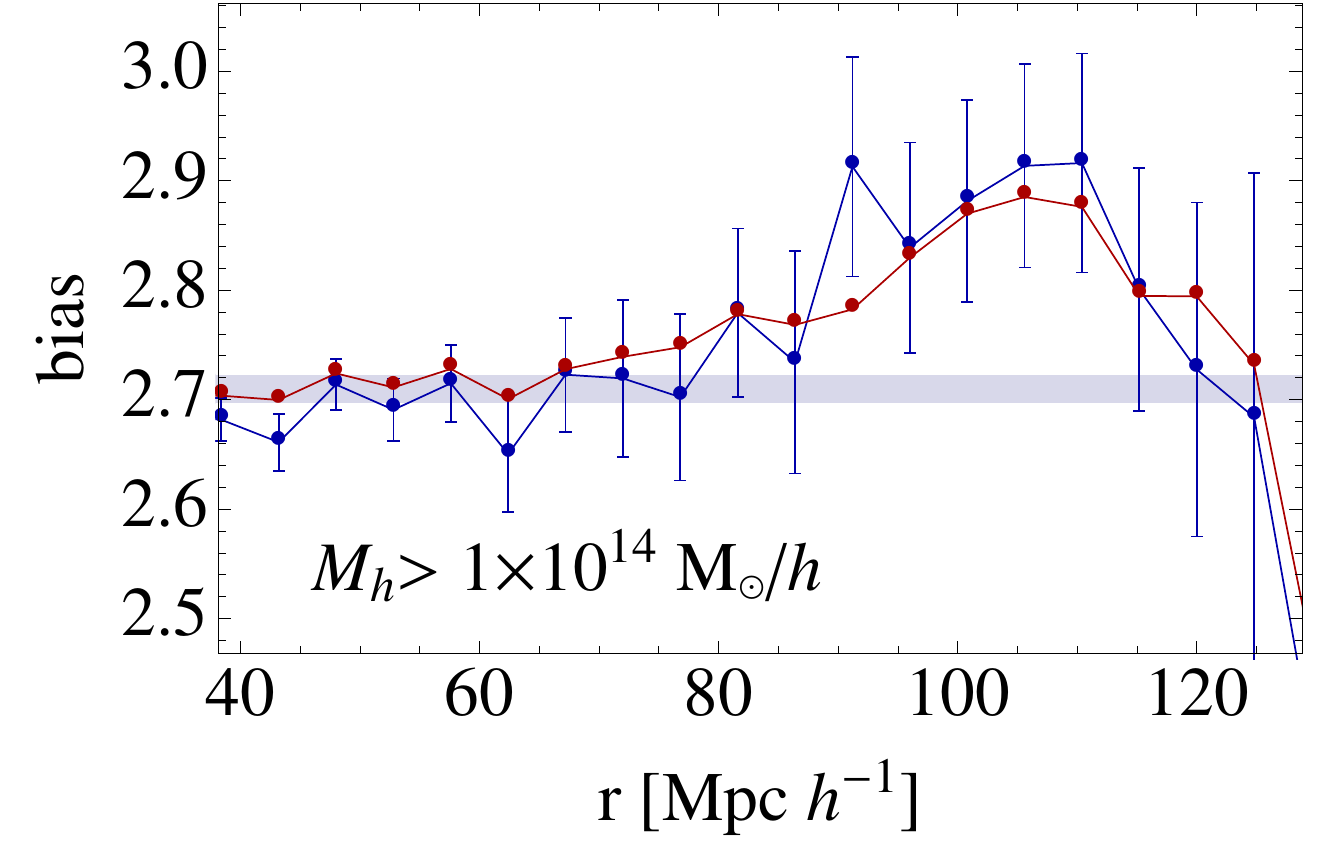} 
\caption{Scale dependence in bias across the BAO feature for a cluster
  mass scale halo sample. The sample is selected from the comoving
  output at $z=0$. Shaded regions as in
  Fig.~\ref{fig:BAObiashalos}. Error bars are only displayed for bias
  derived from halo auto-correlation for clarity, as in Fig.~\ref{fig:BAObiashalos}} 
\label{fig:BAObiasclusters}
\end{center}
\end{figure}

\begin{figure*}
\begin{center}
\includegraphics[trim= 0.8cm 1.2cm 0cm 0cm, clip=true, width=0.8\textwidth]{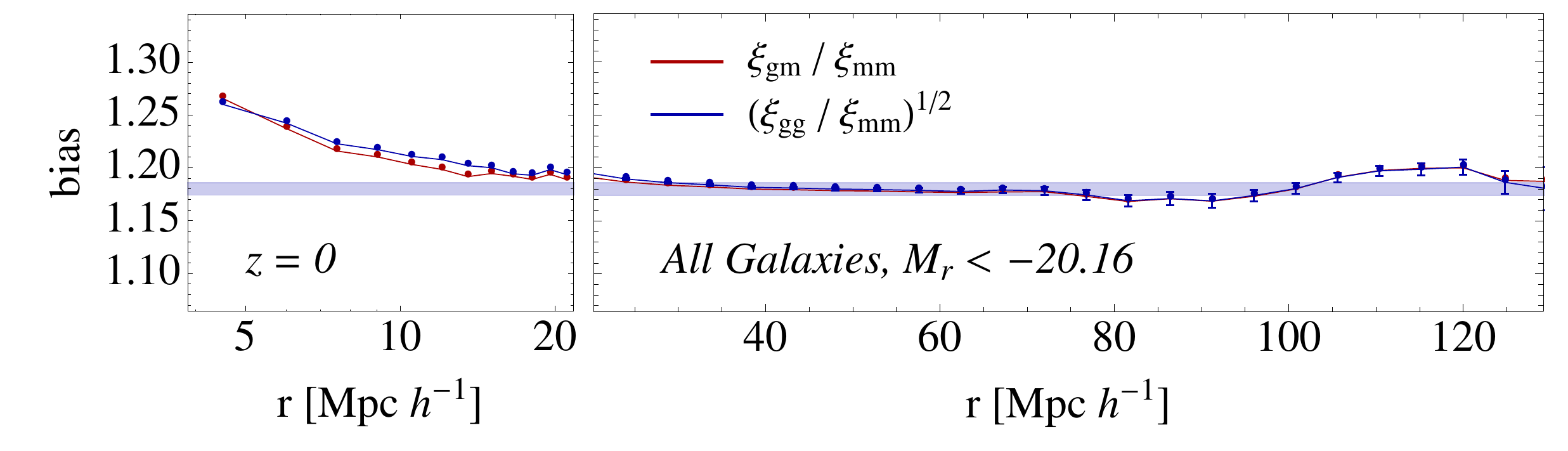} \\
\includegraphics[trim= 0.8cm 0.3cm 0cm 0cm, clip=true, width=0.8\textwidth]{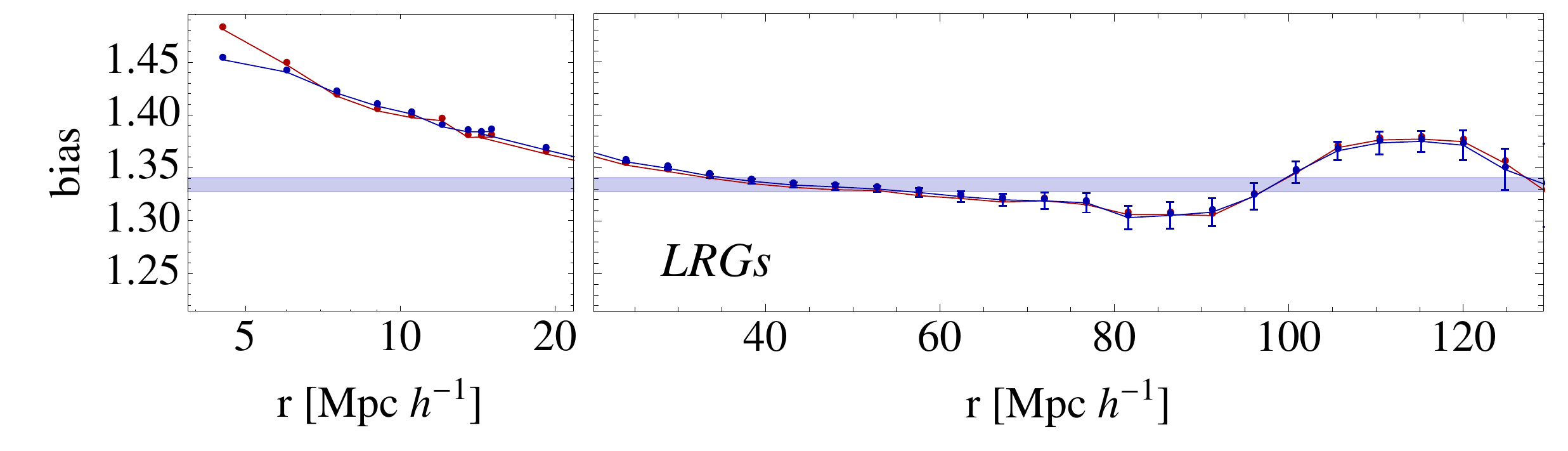} 
\caption{Scale dependence in galaxy bias for two samples in the
  MICE-GC comoving catalogue at $z=0$. Top panel correspond to an absolute
  magnitude limited sample ($M_r<-20.16$). Bottom panel to an LRG-like
  sample ($M_r < -21$ and $g-r > 0.8$). The shaded region correspond
  to $1\%$ around the mean bias for $r>30\Mpc$. The panels show trends resembling those present for halo
  clustering. The $M_r< -20.16$ galaxies show a remarkably flat bias for $r>20\Mpc$ while the LRGs have a scale dependent
  feature across the BAO region of order $4-6\%$. Stronger
  nonlinear effects show up at smaller scales (for $r < 20 \Mpc$) but,
  contrary to the case of halos or centrals only shown in Fig.~\ref{fig:BAObiashalos}, the
  cross correlation coefficient remains close to unity down to
  $\sim {\rm few} \Mpc$ scales}
\label{fig:BAObiasgalaxies}
\end{center}
\end{figure*}

For the least massive halos there seems to be a decrement of clustering
amplitude around $100\Mpc$, although with a marginal amplitude of about $2\%$.
As we increase the sample mass to $M_h \gsim M_{\star}$\footnote{Where
  $M_\star$ is defined as the mass scale with a variance
  $\nu=\delta_c/\sigma(M_\star)=1$, which for MICE yields $M_\star
  =2.3\times 10^{12}\Msun$ at $z=0$.} (see middle panel) the bias is consistent with linear bias almost within $1\%$.
However for masses considerably above $M_\star$ (such as $M_h
\ge7\times10^{10}\Msun$ as shown in the bottom panel of
Fig.~\ref{fig:BAObiashalos}) we find an excess
of clustering of $4\%-5\%$ precisely at the BAO peak ($r \approx 110 \Mpc$ for our
cosmology). This excess clustering increases with increasing mass, for
$M_h \gtrsim 10^{14}\Msun$ is separately shown in Fig.~\ref{fig:BAObiasclusters}. This
pattern has been discussed by \cite{2010PhRvD..82j3529D} in the
context of peak biasing and attributed for the most part to first
order effects in Lagrangian Space, but more work is needed to
characterize this as a function of mass and redshift. 

At scales smaller than $20\Mpc$ the bias becomes steadily scale dependent due to
nonlinear gravitational effects ($b_2$ terms in the language of PT)
, again with an interesting dependence with mass since $b_2$ changes
sign from negative to positive across the three halo samples shown
(top to bottom respectively). Notice that this is the expected behavior for $b_2$ given the
values of the linear bias $b_1$ at large scales \citep{2002PhR...372....1C}.

At these scales the cross-correlation coefficient also departs
from unity by up to $10\%$ at $r=5\Mpc$ (see also
\cite{2013PhRvD..87l3523S}). This is compatible with the emergence of
nonlinear bias on these scales but might also signal a stochastic
relation between halos and mass.

We next turn to investigate similar issues in the galaxy 
catalogue. Fig.~\ref{fig:BAObiasgalaxies} shows the galaxy bias from
galaxy-mass cross correlations (red line) and from galaxy-galaxy
auto-correlations for two distinctive samples. The top panel focuses
on one faint magnitude limited sample ($M_r < -20.16$), already discussed in
Sec.~\ref{sec:halobiaspk}, for which the corresponding halo mass from the HOD+HAM
mass-luminosity relation is $M_h\sim 10^{12}\Msun$. Bottom panel
corresponds to a Luminous Red Galaxy sample defined through a bright
absolute luminosity cut ($M_r < -21$) and one color cut $(g - r) > 0.8$ (rest-frame colors). 
Both samples includes all galaxies, making no distinction between
centrals and satellites, and hence populate in a non-trivial way a range of
halo masses.

The magnitude limited sample shows a bias remarkably close to scale
independent across BAO scales and down to $r \sim 20 \Mpc$, where
nonlinear effects increase the clustering above the linear value.
Notably the cross-correlation coefficients
remains tightly close to unity all the way to $r \sim 5\Mpc$, a clear
and remarkable difference
with respect to the halos in Fig.~\ref{fig:BAObiashalos}. In a follow
up work we will explore to what extent this depends on the satellite
profiles or the halo exclusion. But for instance notice that
our satellite galaxies do not necessarily follow the distribution of matter as
we place them using a pre-determined profile.

Turning to the LRG sample in the bottom panel of
Fig.~\ref{fig:BAObiasgalaxies} we find a clear scale dependent bias across the
BAO feature, with an excess power at the BAO peak of about $5\%$ and a
small $2\%$ deep at $80 < r/(\Mpc) < 100$. A more detailed
characterization of these
effects as well as an interpretation from the theory point of view is left for further work. We
note however that this kind of scale dependent residuals across BAO scales is
relevant for an accurate calibration of the standard ruler test. 
  We also note
that our results are in qualitative agreement with other work in the literature
(e.g. \cite{2014MNRAS.442.2131A,2011ApJ...734...94M,2009PhRvD..80f3508P}).

\section{Redshift Space Distortions} 
\label{sec:rsd}

In this section we will discuss the properties of our galaxy catalogue
in redshift space, which is a measure of how galaxy velocities are assigned.

\subsection{Kaiser Limit and bias in the Lightcone}
\label{sec:rsdgal}

In the large-scale linear regime and in the plane-parallel 
approximation (where galaxies are taken to be sufficiently far away
from the observer that the displacements induced by peculiar 
velocities are effectively parallel), the distortion caused by
coherent  infall velocities takes a particularly simple form 
in Fourier space \citep{kaiser84}:
\begin{equation}
\label{eq:kaiserdelta}\delta^{(s)}(k,\mu) = (1 + f \mu^2) \delta_m(k)
\end{equation}
where $\mu$ is the cosine of the angle between $k$ and 
the line-of-sight, the superscript $s$ indicates redshift space, and
$f(z)$ is given by, 
\beq
f(z) \equiv \frac{d\;ln\;D}{d\;ln\;a}.
\eeq
The second term in Eq.~(\ref{eq:kaiserdelta}) is caused by radial peculiar velocities.
If we assume that galaxy fluctuations
are linearly biased by a factor $b$ relative to the underlying matter density 
$\delta_m$ (i.e. $\delta_G = b \, \delta_m$)  but velocities are unbiased,
then 
\begin{equation}
\label{eq:kaiserdeltag}\delta^{(s)}_G(k,\mu) = (b + f \mu^2) \delta_m(k)
\end{equation}
where $\delta^{(s)}_G$ are the measured galaxy fluctuations in redshift
space. We then have an anisotropic power spectrum:
\beq
P^{(s)}_{gg}(k,\mu) = <(\delta^{(s)}_G(k))^2>= (b+ f \mu^2)^2 P_{mm}(k) 
\label{eq:kaiserpk}
\eeq
where $P_{mm}(k)=<\delta^2_m(k)>$ is the real space matter power spectrum.
This can be Fourier transformed and averaged over angles to obtain the monopole correlation function:
\begin{eqnarray}
\xi_{gg} &\equiv& \xi_{\ell=0,gg} =  K_{\ell=0}(z)  ~\xi_{mm} \nonumber \\ 
\indent K_{\ell=0}(z) &\equiv& b(z)^2+\frac{2}{3} b(z) f(z)+\frac{1}{5} f(z)^2b(z)^2
\label{eq:xiKai}
\end{eqnarray}
where $\xi_{mm}$ is the matter
correlation function at redshift $z$ (i.e. in linear theory
$\xi_{mm} = D^2(z) \xi_{\rm L}(r,z=0)$)
and we have defined $K_0(z)$ to be the monopole ``linear Kaiser'' factor.

\begin{figure}
\begin{center}
\includegraphics[trim= 0cm 0.5cm 0cm 1cm, clip=true, width=0.44\textwidth]{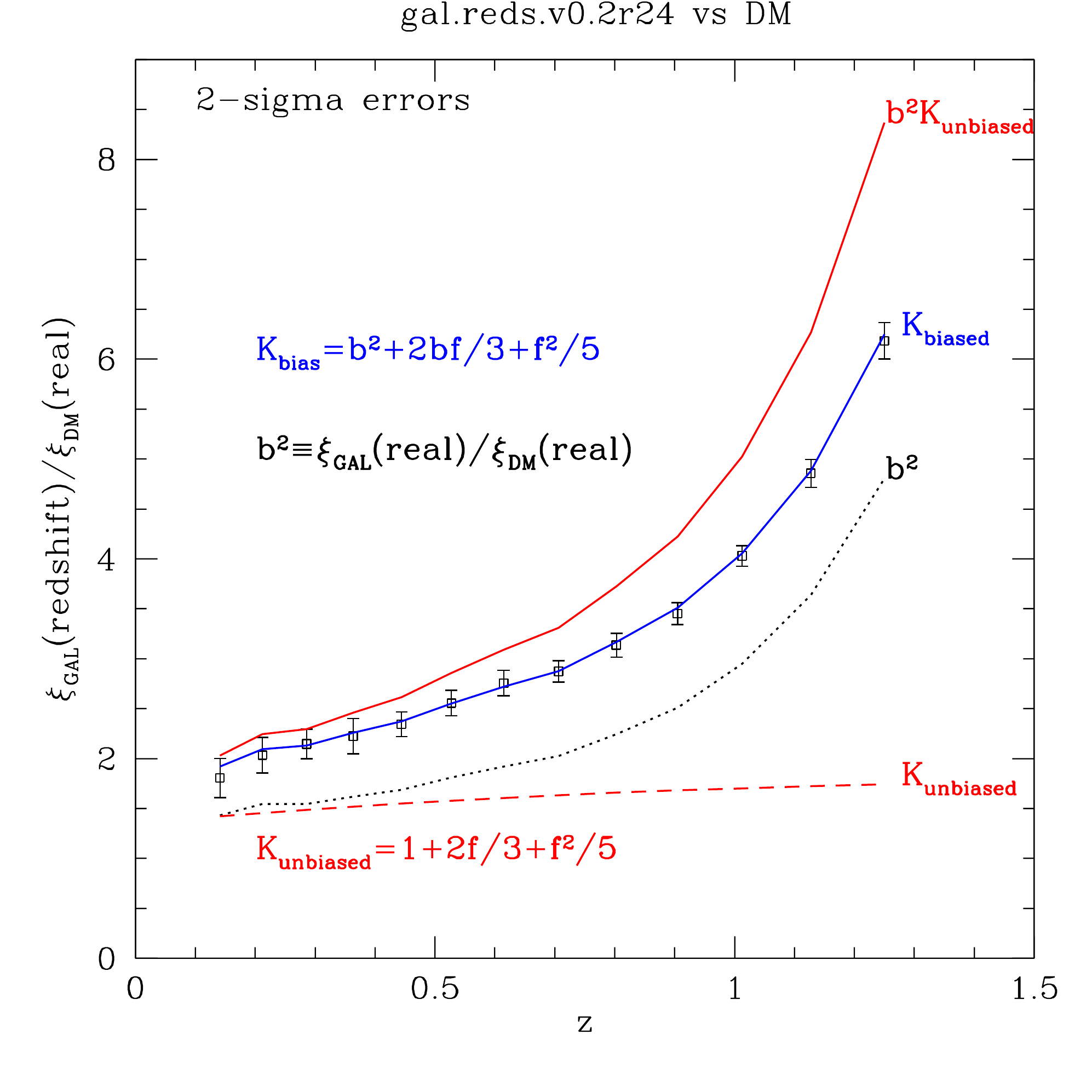}
\caption{Ratio of galaxy monopole 3D correlations in redshift space to the
matter correlation in real space (points with errors), see Eq.~(\ref{eq:xiKai}). Dashed line
shows the unbiased Kaiser prediction, while dotted line shows
the bias measured in real space averaging over scales $s \ge 30\Mpc$. The blue line corresponds
to the linear Kaiser model in Eq.~(\ref{eq:xiKai}) with this
measured bias. This correspond to
$r<24$ galaxies in the MICE-GC lightcone catalogue.}
\label{fig:Kai2}
\end{center}
\end{figure}

Figure~\ref{fig:Kai2}  shows the ratio $\xi_{gg}/\xi_{mm}$ measured in the
MICE-GC galaxy lightcone catalogue 
(error weighted averaged on scales $r>30 \kvecMpc$)
for an apparent magnitude limited sample ($r<24$). We compare it 
to the linear Kaiser factor $K$, where we use $b(z)$ as measured in real
space and $f(z)$ given by the MICE cosmology. Note how
both $b(z)$ and $f(z)$ change with redshift and that the predictions
depend strongly on both ($b$ or $f$ alone cannot account for the observed
variations, as indicated by red and dotted lines).
There is an excellent agreement with 
the linear Kaiser model (in blue) for all redshifts and for
the concrete bias evolution
that results from cutting galaxies to $r<24$.
On the one hand this serves as an excellent validation of the
large-scale bulk galaxy velocities
in the catalogue (which in turn are based on halo velocities).
On the other hand, it means that both bias $b(z)$ and $f(z)$ can be constrained from observations
using this simple modeling.

\subsection{RSD for Central and Satellite Galaxies}
\label{sec:rsdz0}

In the previous section we showed that the Kaiser limit is a good
model to describe the large-scale ($s \ge 30\Mpc$) clustering
amplitude of the monopole correlation function and its lightcone
evolution, provided with the bias as a function of redshift. 
In this section we investigate the break down of this large-scale
limit due to the departure from purely bulk motions, in particular the
impact of satellite galaxies.

We will focus on the multipole moments of the anisotropic
galaxy power spectrum in redshift space,
\beq
P^{(s)}_{gg,\ell}(k) = \frac{2\ell+1}{2} \int_{-1}^{1} P^{(s)}_{gg}(k,\mu)
L_\ell(\mu) d\mu
\eeq
with $L_\ell$ being the Legendre polynomials.
On large scales we can assume the ``linear'' relation
in Eq.~(\ref{eq:kaiserpk}) and obtain the corresponding Kaiser limits :
$P^{(s)}_{gg,\ell}(k)= K_{\ell}(b,f) P_{mm}(k)$, where $K_0$ is given in Eq.~(\ref{eq:xiKai})
and,
\begin{eqnarray}
\indent \indent K_2 (b,f)&\equiv& \frac{4}{3} b f + \frac{4}{7} f^2 \nonumber \\
K_4 (b,f)&\equiv& \frac{8}{35} f^2
\end{eqnarray}

\begin{figure}
\begin{center}
\includegraphics[trim= 0.cm 1.1cm 0cm 0cm, clip=true, width=0.45\textwidth]{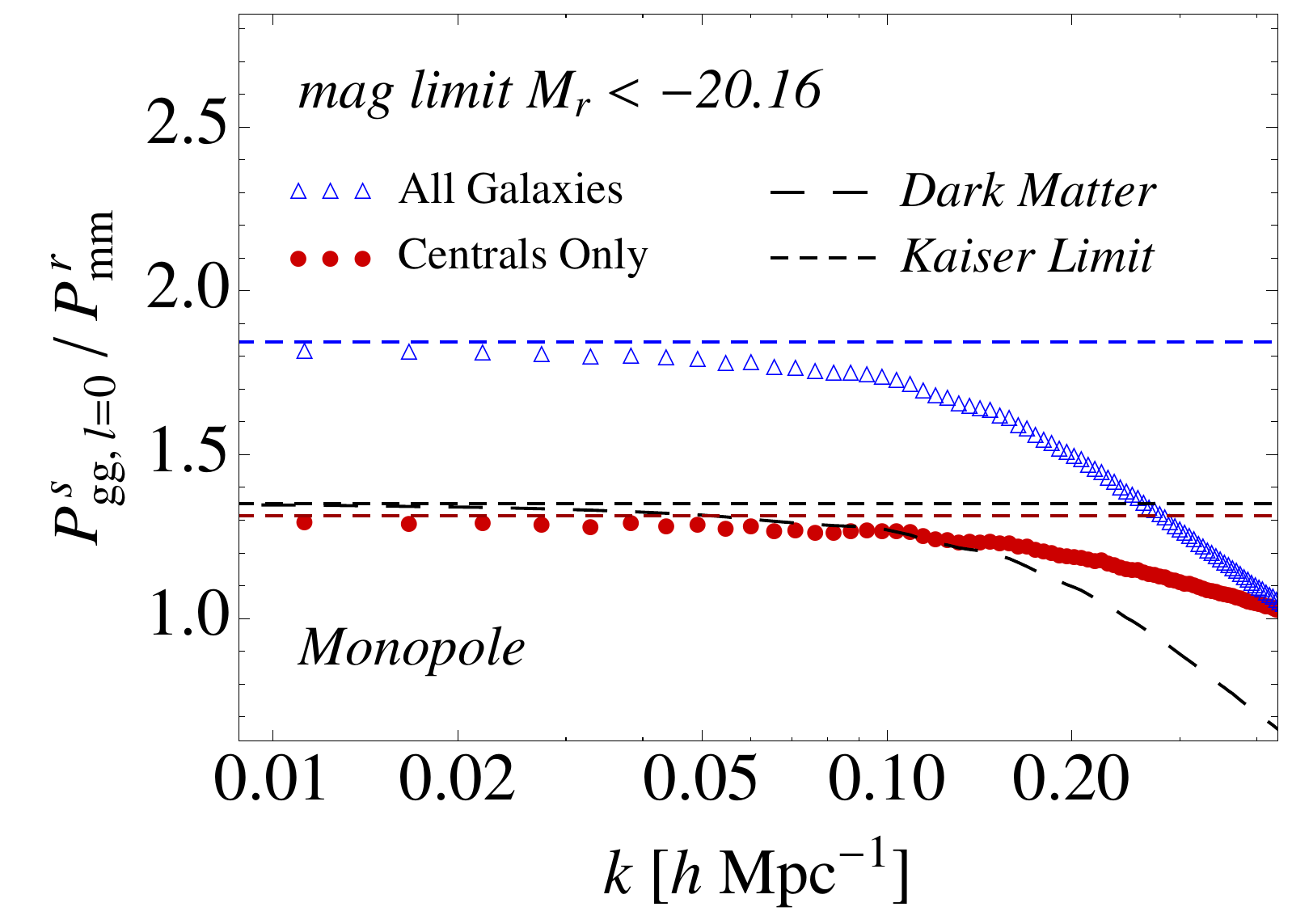} \\
\includegraphics[trim= 0.cm 1.1cm 0cm 0cm, clip=true, width=0.45\textwidth]{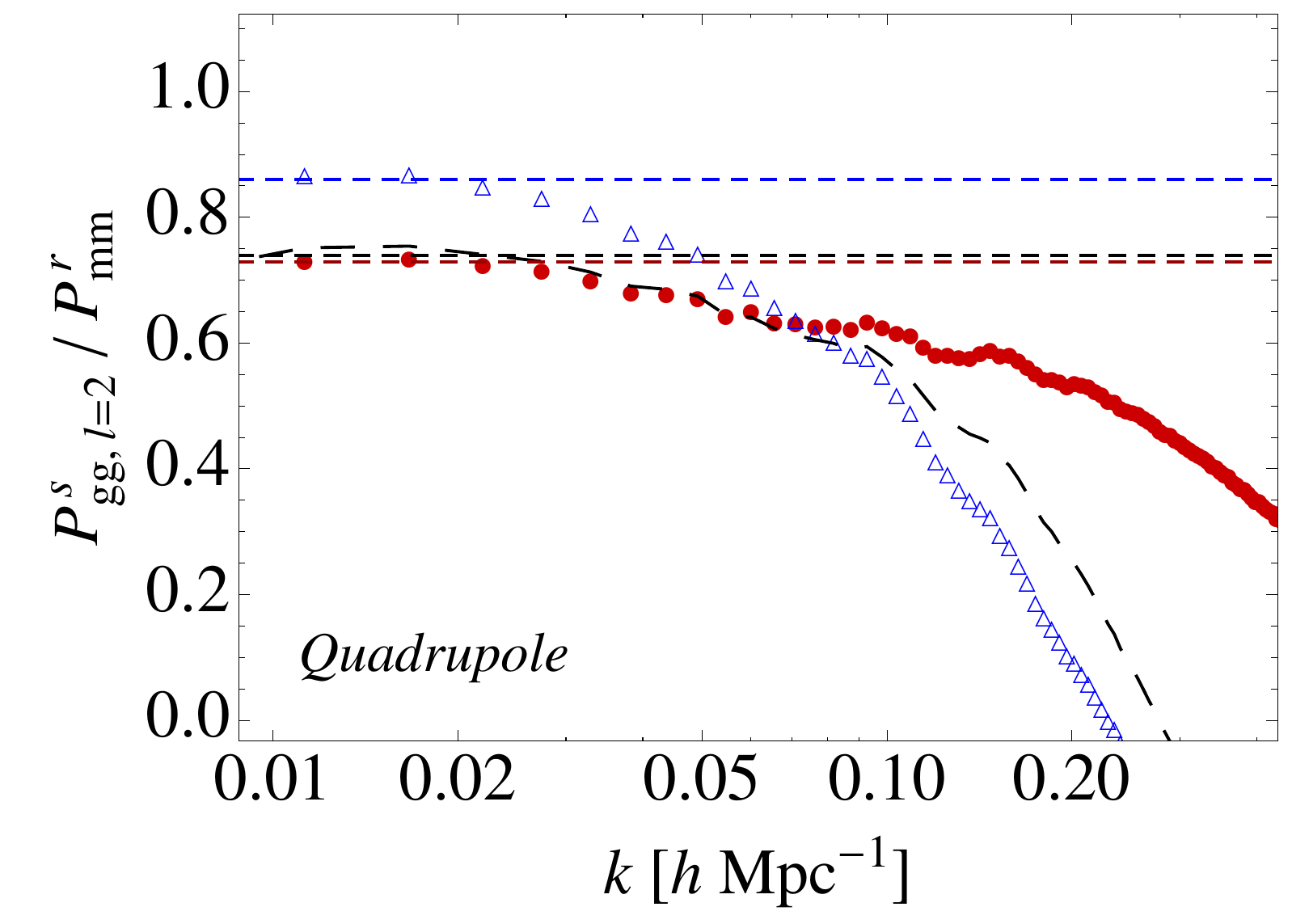} \\
\includegraphics[trim= 0.cm 0.cm 0cm 0cm, clip=true, width=0.45\textwidth]{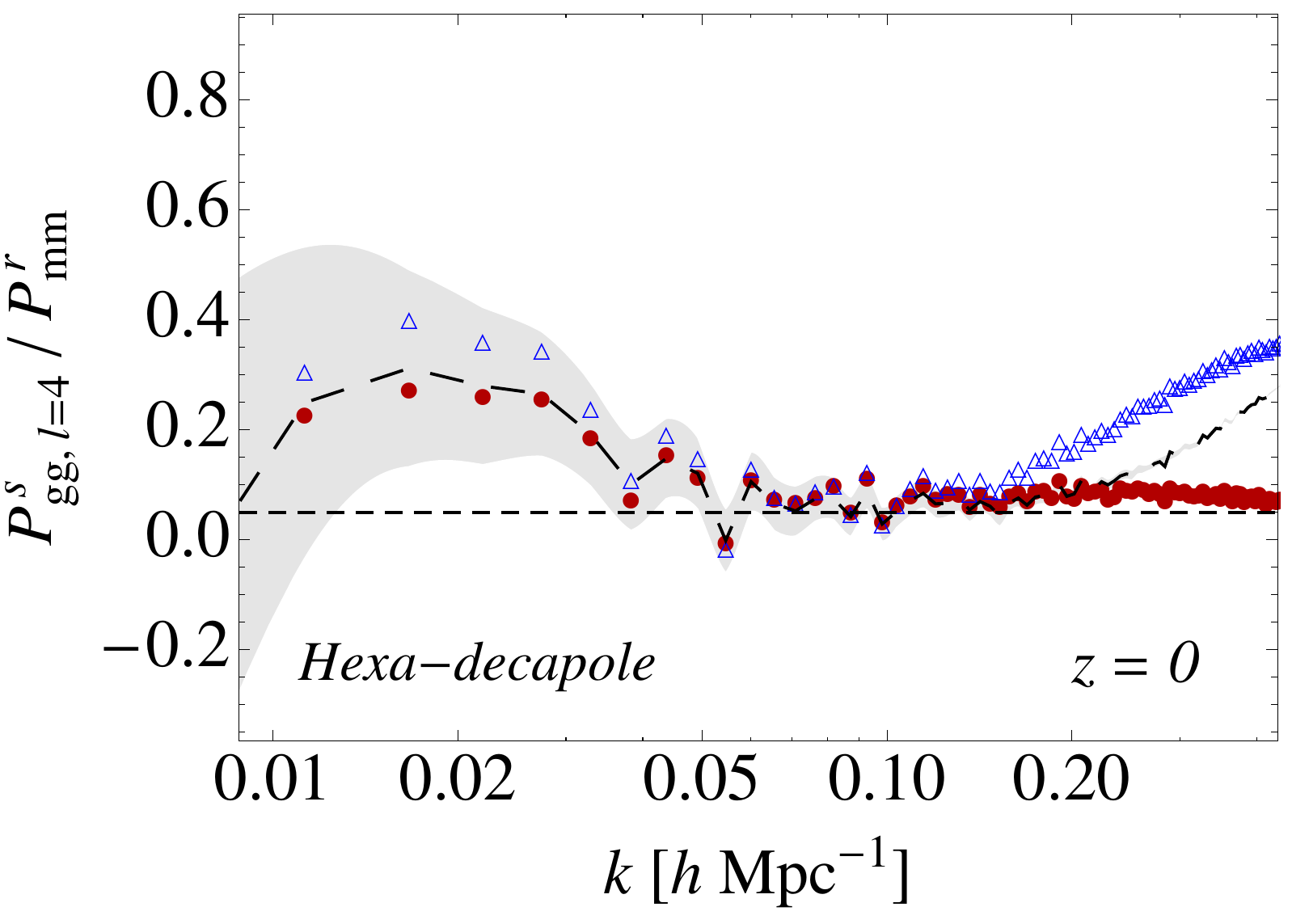} 
\caption{The first 3 multipole power spectra for a magnitude limited
  galaxy sample ($M_r<-20.16$) at $z=0$. In each panel the corresponding $P_\ell$ has
been divided by the measured (non-linear) matter power spectrum. The figure shows the case for central galaxies only, or
the full sample (cen+sat) as well as the corresponding
dark-matter. Hence it stresses the significant impact of satellite galaxies
into the anisotropic clustering, basically by adding velocity
dispersion. Notice how in all cases the Kaiser limit
(short dashed line) is reached but only for the largest scales.} 
\label{fig:rsdsat}
\end{center}
\end{figure}

Figure~\ref{fig:rsdsat} shows the first three moments, monopole ($\ell=0$),
quadrupole ($\ell=2$) and hexadecapole ($\ell=4$) for the 
magnitude limited sample of galaxies discussed in Sec.~\ref{sec:galhaloclustering}, i.e. $M_r <
-20.16$, in the comoving catalogue at $z=0$. In the mean, this luminosity corresponds to halos more
massive than $10^{12}\Msun$. In order to understand what is the impact
in the anisotropy of large-scale fluctuations from the motion of satellite
galaxies inside halos we split the sample into ``centrals only''
(i.e. bulk motion of halos only), shown by red filled dots, and ``all
galaxies'' (central plus satellites) shown by blue empty triangles.
For this sample the satellite fraction is $24\%$.
The corresponding multipole spectra for dark-matter is shown by dashed black lines.

On the largest scales the Kaiser limit (shown in short-dashed) is reached for both the
``centrals only'' sample and the ``central+satellites'', although 
in a more limited range of scales for the later. 
Notice
that the large-scale bias of these two samples is slightly different
($b_{cen}=0.98$ and $b_{cen+sat}=1.2$) because of the scatter in the mass-luminosity relation discussed in
Sec.~\ref{sec:galhaloclustering}. Hence the different Kaiser asymptotics in the monopole and
quadrupole panels of Fig.~\ref{fig:rsdsat} (while $K_4$ does not depend on bias). 
In turn, at $k<0.05\kvecMpc$ sampling variance dominates the hexadecapole
results, despite the large simulation size. For reference we show
the corresponding cosmic variance error
assuming the multipole moments in
redshift space to be Gaussian random fields
(e.g. \cite{2009PhRvD..80l3503T} and references therein).

This is in contrast
to the monopole or quadrupole, which can be measured to
much smaller $k$, and results from
the stronger dependence in the shape ($\mu^4$). 

In order to investigate departures from the Kaiser limit we fit the 
following model to our monopole and
quadrupole measurements \citep{2004PhRvD..70h3007S},
\beq
P^{(s)}(k,\mu)=
\left[b^2 P_{\delta\delta} +2 b f \mu^2 P_{\delta\theta} + f^2
  \mu^4 P_{\theta\theta} \right] \times {\rm e}^{-(k \mu f \sigma_v)^2}
\label{eq:rsdmodel}
\eeq
where we take $b$ to be the large-scale linear bias measured in real
space (i.e. with a fixed value),
$f=0.46$ for our cosmology at $z=0$ and $\sigma_v$ is a nuisance parameter
related to (1D) velocity dispersion. In Eq.~(\ref{eq:rsdmodel}) $P_{XY}$ are the
nonlinear density ($\delta$) and velocity divergence ($\theta$)
auto and cross power spectra which we compute using {\tt MPTbreeze} \citep{2012MNRAS.427.2537C}.
We stress that Eq.~(\ref{eq:rsdmodel}) is not expected to give accurate results
but it is useful to hint on departures from the simplest linear Kaiser
model discussed before.

From the monopole and quadrupole in Fig.~\ref{fig:rsdsat} we
find the best-fit\footnote{We limit to scales
  $k\le 0.13\Mpc$ where the model fits the three multipoles,
  provided with one nuisance parameter.} to be
$\sigma_v=6\Mpc$ for dark matter (equivalent to $600\,{\rm km/s}$), very close to the linear value 
\beq
\sigma_{v, \rm Lin}= \left( \frac{4\pi}{3}\int P_{Lin}(q) dq
\right)^{1/2} = 6.15\Mpc,
\eeq
in agreement with \cite{2010PhRvD..82f3522T}. The ``centrals'' only
sample (or halos) yields a smaller value
$\sigma_v^{cen}=3\Mpc$ characteristic of a more coherent 
 bulk motion. In turn the inclusion of satellite
galaxies leads to virialized motions closer to those of dark matter,
with a best-fit $\sigma_v^{all}=8.5\Mpc$. 


\subsection{Changing the velocity dispersion of satellite galaxies}
\label{sec:satveldisp} 

As we discussed in Sec.~\ref{sec:hod} the spatial distribution of satellite galaxies is
set by observational constraints from projected clustering
\citep{carretero2014}. However equivalent observational constraints for 
the distribution of satellite velocities are not that well stablished, 
hence our choice arises from well known results using hydrodynamical
simulations \citep{1998ApJ...495...80B}. In this section we study
quantitatively how this assumption
impact the anisotropic clustering. 

Our procedure is to give the satellite galaxies the bulk motion of the
halo plus an additional virial motion that follows a Gaussian
distribution (in each axis) with a velocity dispersion $\sigma_{vir}
= \langle v_{vir}^2 \rangle \propto M^{2/3}_h$ \citep{2001MNRAS.322..901S}, so
$v=v_h+v_{vir}$  where $v_h$ is the halo center of mass velocity (also the
one of the central galaxy). 

\begin{figure}
\begin{center}
\includegraphics[trim= 0.cm 1.1cm 0cm 0cm, clip=true, width=0.45\textwidth]{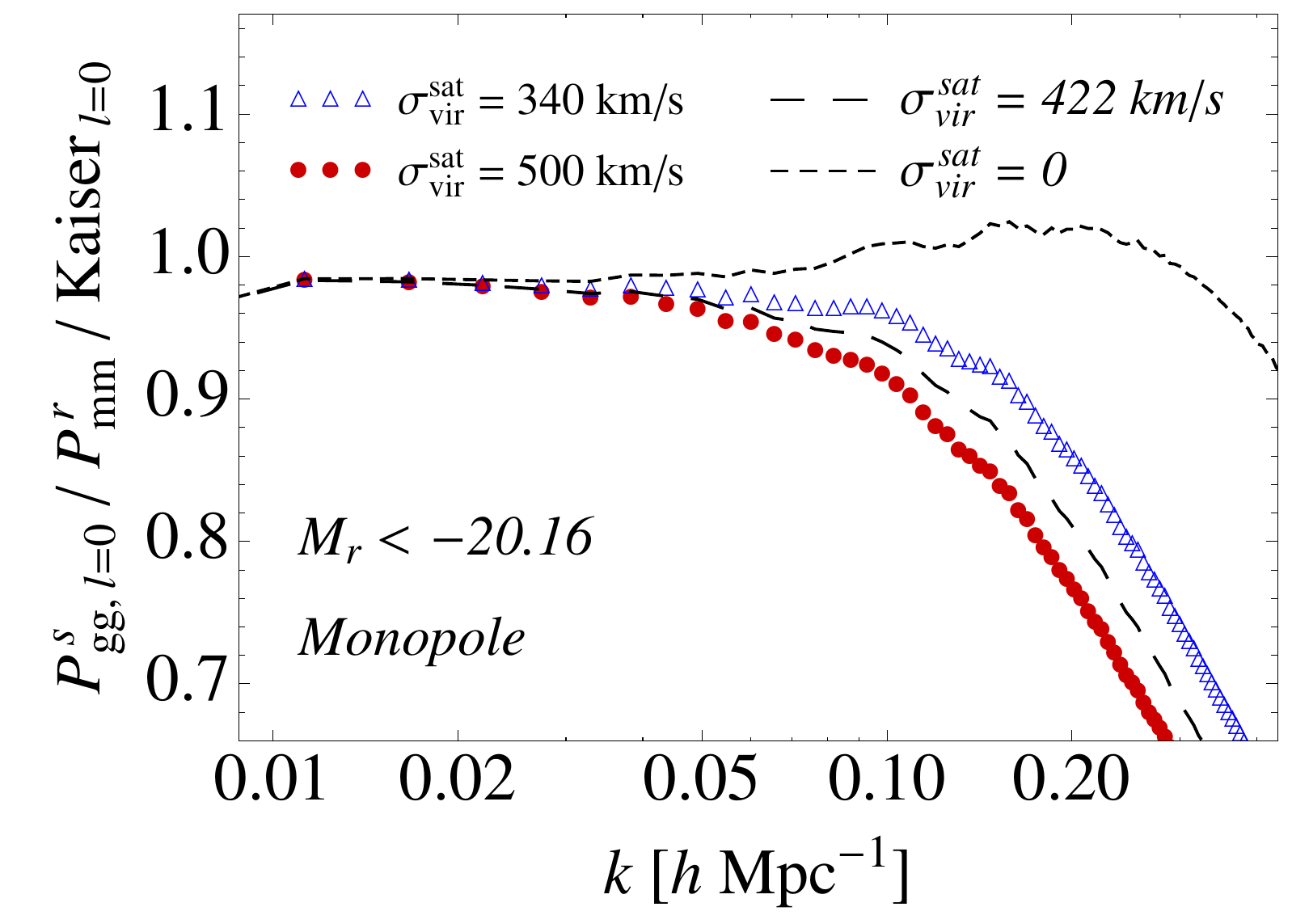} \\
\includegraphics[trim= 0.cm 1.1cm 0cm 0cm, clip=true, width=0.45\textwidth]{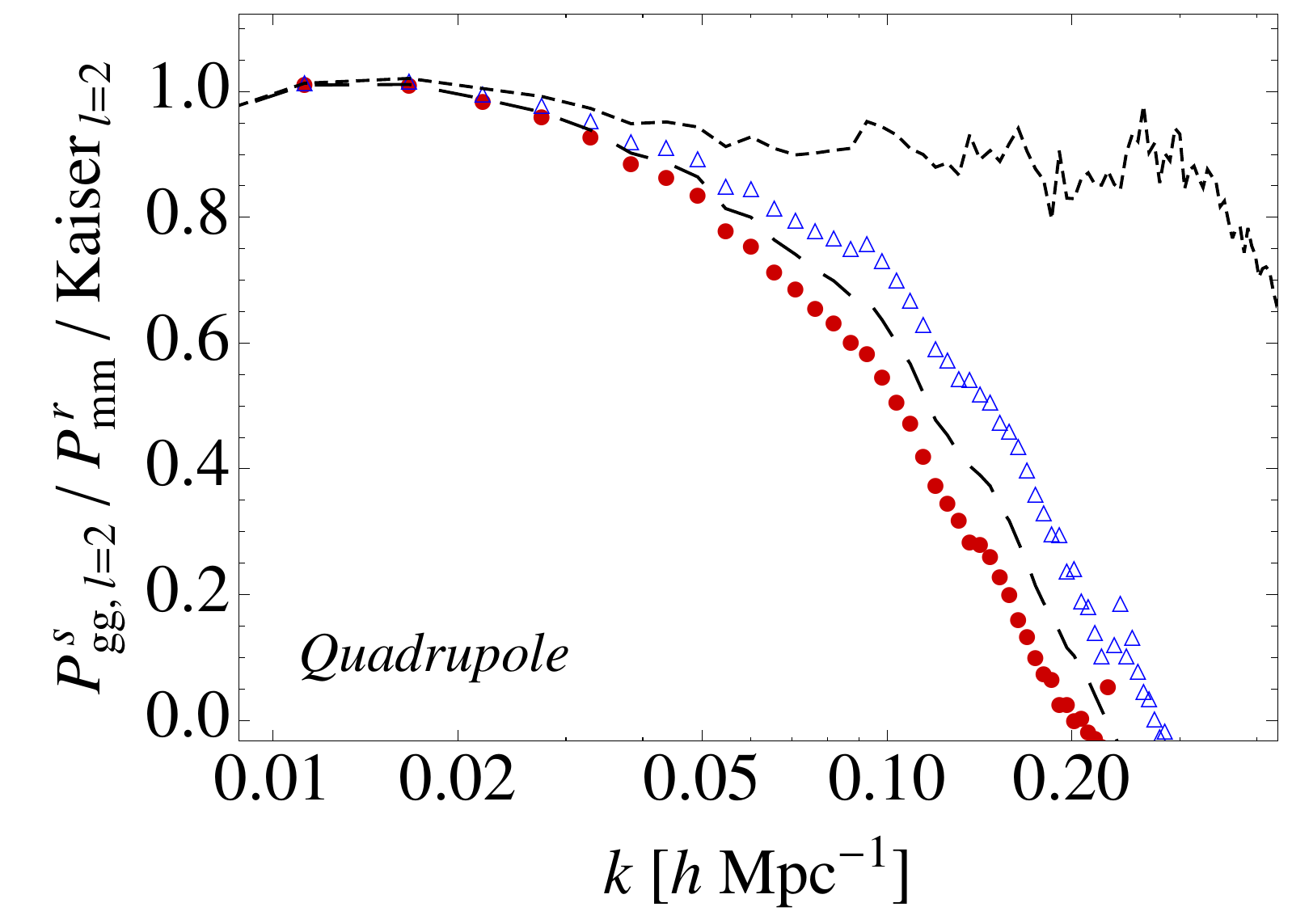} \\
\includegraphics[trim= 0.cm 0.cm 0cm 0cm, clip=true, width=0.45\textwidth]{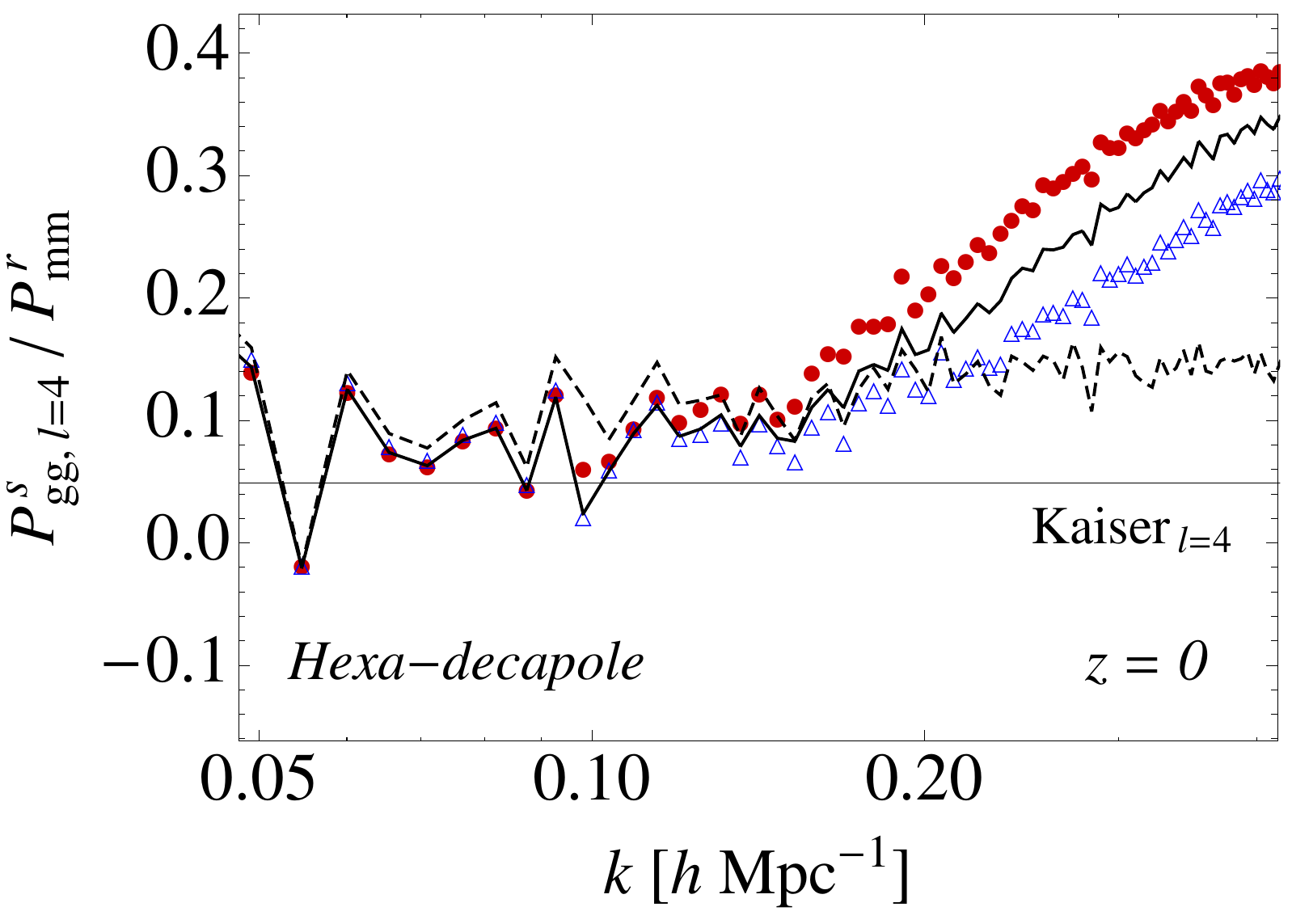} 
\caption{Change in the first 3 multipole power spectra for a magnitude limited
  galaxy sample ($M_r<-20.16$) when the virialized motion of satellite
galaxies within the host halos is changed by $\sim \pm 20\%$. We
normalized the measurements by the corresponding Kaiser prediction and
the measured dark matter spectrum. The short dashed line corresponds
to the extreme case where all satellite galaxies move with the bulk
motion of their host halos.} 
\label{fig:satveldisp}
\end{center}
\end{figure}

For the magnitude limited sample $M_r<-20.16$ discussed in
 Sec.~\ref{sec:rsd} the velocity dispersion of satellites is $422 \, {\rm Km \,sec^{-1}}$ (the distribution is narrower than a
Gaussian, because it arises from a range of halo masses) while the
satellite fraction is $\sim 30\%$. In
Fig.~\ref{fig:satveldisp} we show how the monopole, quadrupole and hexadecapole
change when the satellite velocity dispersion is changed by $\pm 20\%$
keeping the bulk motion of the halos unaltered. As expected increasing
the satellite velocity dispersion to $500\,{\rm km/s}$ induces more FoG
effects (from satellite-central correlations in different halos) and a
stronger scale dependent suppression of power.
The monopole is suppressed at the $5\%$ level on scales $k \sim 0.15
\kvecMpc$ 
compared to the fiducial case, while the quadrupole is more
affected ($20 \%$ at the same scale). In turn the hexa-decapole
is too noisy on these scales, but the impact is clearly
stronger. Reducing the satellite velocity dispersion to $340\,{\rm km/sec}$
(i.e. by $20\%$ less w.r.t the fiducial) has the opposite effects.  We have
done a more extreme case in which all satellites move with the bulk
motion of the halo (setting $v_{vir}=0$). This is shown by
short-dashed lines in Fig.~\ref{fig:satveldisp}. The result is that the anisotropic
clustering in this case is well described by the simple linear Kaiser
effect down to smaller scales.

Overall we find that satellite galaxies give a significant contribution to
the anisotropic clustering through nonlinear redshift space
distortions even on quite large-scales (see also
\cite{2013JCAP...08..019H,2013MNRAS.433.3506M,2014MNRAS.444.1400N}
for the case of LRG's), yielding velocity dispersion effects similar (or larger) to those of dark-matter.

\section{Public Galaxy Catalogue Release}   
\label{sec:release}   

Together with this series of papers we make a first public data release of
the current version of the MICE-GC lightcone catalogue ({\tt MICECAT v1.0}). The
halo and galaxy catalogue can be obtained at {\tt
  http://cosmohub.pic.es}, a dedicated database portal hosted by 
Port d'Informaci\'o Cient\'ifica (PIC). 
It corresponds to one octant
of the full sky ($5000\,{\rm deg}^2$) from $z=0$ to
$z=1.4$. In the current version (v1.0) galaxies are limited in absolute magnitude to $M_r < -18.9$
and halos in mass to $M_h > 2.2\times 10^{11} \Msun$. Among other
properties we provide
angular positions and observed redshifts for all galaxies, flags to
central/satellite distinctions, host halo masses, peculiar velocities,
comoving distances, observed magnitudes (and magnitude errors) in
several bands for surveys such as SDSS, DES and VISTA. Besides, we provide
the SED template and dust extinction assigned to each galaxy which rely
on the template library of \cite{2009ApJ...690.1236I}. These SEDs were
used to generate the observed magnitudes. We also
provide lensing information for each galaxy such as shear and
convergence, as well as magnified magnitudes and angular positions
(described in Paper III). Lastly we provide photometric redshift
errors and error distributions based on a photometric  template code. 

The web-portal is
set up to facilitate the download of the data, either the full compressed
catalogue or by querying particular regions of the sky or data columns.

\section{Conclusions}   
\label{sec:conclusions}   

We have presented the MICE-GC halo and galaxy catalogues
built upon one the largest N-body runs completed to date: the MICE Grand Challenge
lightcone simulation. 
This N-body run contains about 70 billion particles in a 3 $\Gpc$
periodic box, a unique combination of large volume and fine mass
resolution sampling 5 orders of magnitude in dynamical range.

We identify bound structures using a Friends-of-Friends algorithm
with linking length $b=0.2$. Halos were resolved down to few times
$10^{11}\Msun$ with a total of about $157$ million identified in
each octant of the full sky
lightcone up to $z=1.4$. A similar procedure was followed in several
comoving outputs.
We then populated the dark-matter halos
with galaxies following a hybrid
HOD and HAM
scheme,  matching the luminosity, color distributions and projected clustering properties (as a
function of luminosity and g-r color) of SDSS galaxies at low-$z$. Lastly
galaxy properties were evolved into the past lightcone using stellar
evolution models. In all, this resulted in a catalogue limited in absolute
magnitude to $M_r<-18.9$ and containing $\sim 2\times 10^8$ galaxies 
(considering only one octant of the full sky and $z<1.4$).

We have performed several validation tests of the catalogues, with the following main conclusions,

\begin{itemize}

\item{
{\it Halo Catalogue: } We showed that the halo mass function at $z=0$ agrees at
the $1\%-2\%$ level with the \cite{crocce10} fit for well resolved halos 
(similarly for other comoving redshifts and the lightcone, were the
fit does not assume universality). The cumulative
  abundance of groups with as low
as 10 particles is up to $15\%$ below the model prediction using the
\cite{crocce10} fit (which is a numerical fit calibrated to higher-resolution runs).
In turn, the MICE-GC resolution and volume allow us to
study halo clustering with good precision for samples with a broad range of
linear bias values, even $b\lesssim1$. The PBS prediction for this low bias
sample agrees at the $2\%$ level with the bias measured from
$P_{hm}/P_{mm}$, a better
performance than for massive objects \citep{2010MNRAS.402..589M}. We note that this regime of
low-bias was not well explored previously and deserves a more detailed
analysis for more robust conclusions about the
  performance of PBS. Lastly, halos in the lightcone presented an almost
constant clustering amplitude, i.e. degenerate with the growth factor
evolution, for constant mass samples. Instead, galaxies selected above
an apparent luminosity threshold show a clustering amplitude that
increases with redshift.}\newline

\item{
{\it Galaxy Catalogue: }Starting from fits at low redshift and
implementing evolutionary 
corrections to galaxies and resampling SEDs
from COSMOS \citep{2009ApJ...690.1236I}, we were able to predict the color distributions and clustering properties of
higher-redshift galaxy populations ($0.8<z<1.4$). In particular, MICE
and COSMOS galaxies have very similar color-color distributions at
low-$z$, whereas the MICE sample is slightly bluer at higher redshifts,
as depicted by Fig.~\ref{fig:gr-ri}.
As for the clustering, MICE mock galaxies match
very well the shape of the  angular correlation function of  COSMOS
galaxies at $z=0.6$ for a sample brighter than $i_{AB} <
22.5$. A similar match is found at $z=1$ for galaxies with $i_{AB}<24$, except for the rise in clustering strength in COSMOS at 
angular scales larger than 5 arc minutes, which we attribute to the
known excess of clustering power in the COSMOS field \citep{skibba13}.
Compared to the dark matter the galaxy clustering of these samples
is consistent with a simple linear bias model with $b\sim 1.16$
and $b\sim1.8$ respectively, for scales $\theta \gtrsim$ 1 arc minute (see
Figs.~\ref{fig:COSMOS_MICE_z10_wtheta} and
\ref{fig:COSMOS_MICE_z75_wtheta}). We also built a sample resembling
the DR10-BOSS CMASS sample of LRGs which implied doing a magnitude and
color selection over MICE-GC galaxies. The resulting redshift
distribution and clustering (monopole and quadrupole) were in good
agreement with the corresponding measurements in DR10.}
\end{itemize}
We have then explored some concrete applications for these catalogues.
The main findings in this regard are as follows:
\begin{itemize}

\item{
We have studied how the large-scale halo clustering depends on the
 mass resolution of the underlying N-body simulation. 
We focused first in the halo-matter cross-power spectrum which is a robust
measure of halo clustering against shot noise. Using this
estimator we find the bias to be up to $5\%$ larger
for halos resolved with $20-50$ particles in our MICE-IR run than for
the corresponding sample in MICE-GC 
 (a factor of 8 more particles), and $10\%$ for $10-20$ particle halos. 
 The exact value depends on whether halo
 masses are corrected for discreteness following \cite{2006ApJ...646..881W} or not (for poorly
 resolved halos the applicability of this correction is unclear and makes the effect
 worse). For well resolved halos we find no significant difference between
 MICE-IR and MICE-GC large scale clustering.                            
 Although we concentrated in the comoving output at $z=0.5$ we have reached similar conclusions at $z=0$.
}\newline

\item{We also looked into this effect in higher order
    statistics by measuring the reduced 3-point function, $Q_3$, of
    massive halos $M_{h} \ge 1.83\times 10^{13}\Msun$ (sampled with $N_p \ge 625$ in MICE-GC,
    $N_p \ge 78$
    particles in MICE-IR and only $5$ particles in MICE-SHV) at $z=0$.
    Mass resolution effects for this halo resolution do not affect the shape of
    the 3-point function unless we use extremely low resolution as for MICE-SHV.
    Although the MICE-SHV halos yield the correct shape for $Q_3$ there are few
    percent level differences. For smaller scales MICE-GC and MICE-IR
    deviate clearly from the simple local model and track well the
    non-local prediction from \cite{2012PhRvD..85h3509C}, see Fig.~\ref{fig:q3hr96}. 
  }\newline

\item{We investigated scale dependent bias from small (few $\Mpc$) to
    large BAO scales (up to $\sim 130 \Mpc$) 
    in the two-point correlation function of halos and galaxies at $z=0$. We focused on three halo mass threshold samples, $M_h/(\Msun)
  \ge 5\times10^{11},5\times10^{12}$ and $7\times10^{13}$, and found
  the bias to be remarkably close to scale independent (within $2\%$) for scales $ 20 \lesssim
  r/(\Mpc) \lesssim 80$. For the intermediate mass scale (roughly
  $M_\star$ halos) the bias is
  flat also across the BAO. However for more massive halos we find an
  excess of clustering at BAO scales of $\sim 5\%$, while for less
  massive objects we instead find an almost marginal decrement of
  clustering amplitude of $\sim 3\%-4\%$ at $80 \lesssim r/(\Mpc) \lesssim 110$. Stronger nonlinear effects show
  up at scales of $r<20\Mpc$ together with departures of the cross-correlation
  coefficient $r_{cc}$ from unity. We then investigated how this translates to the
  clustering of galaxies, which as a non-trivial
  combination of the one of halos through the HOD.
  For a faint luminosity cut $M_r<20.16$, corresponding to an
  $L_{\rm cen}=L_{\rm cen}(M_h)$ relation of $M_h\sim
  10^{12}\Msun$, we find the bias to be constant with scale for
  $r>20\Mpc$. In turn, for an LRG type selection (bright $M_r<-21$ and
  red $g-r>0.8$ galaxies) we again find a non-trivial scale dependent bias across the BAO feature
  of about $6\%-8\%$. For galaxies, we find the cross-correlation
  coefficient close to unity down to few $\Mpc$. Overall these are relevant conclusions for
  standard ruler tests that aims to extract distance-redshift
  relations from galaxy clustering (as they impact the observed BAO
  feature) or for modeling the full-shape of the
  correlation function. We leave a more detailed analysis for
  follow-up work.}\newline

{\item Lastly we studied galaxy clustering in redshift space, a testing
  ground for galaxy peculiar velocities. Using the lightcone we find the averaged amplitude of the
  monopole correlation function on scales $r > 30\Mpc$
  to be very consistent with the linear Kaiser model (with an
  input bias from real-space measurements). This was true across all redshifts
  sampled in the lightcone ($z<1.4$) which is a non-trivial test of
  both $b(z)$ and $f(z)$. We next looked into departures from the linear
Kaiser model in the multipole moments of the galaxy anisotropic power
spectrum at the $z=0$ snapshot. While on large scales all multipoles agree with the
Kaiser limit there are departures already at $k\sim 0.05\kvecMpc$. Notably the satellite
galaxies make the anisotropic clustering stronger, in the sense of increasing
Finger-of-God effects to reach (or surpass) those of dark-matter.}\newline

In a series of three papers we introduce in detail the MICE-GC mock
galaxy catalogue, the ending product of an elaborated step-by-step
process that puts together dark-matter, halos, galaxies and lensing,
with a strong observational angle. The success of the largest ongoing
and future cosmological surveys is based upon our ability to develop
suitable 
simulations for their analysis and science.
We make our catalogue publicly available,
with the aim of contributing to the community wide effort in shaping the
upcoming era of precision cosmology.

 \end{itemize}

\section*{Acknowledgments} 
We would like to thank Ramin Skibba and Ravi Sheth for sharing their
experience with HOD implementations. Carlton Baugh for very valuable
insight throughout the project. Ariel Sanchez for his help with the 
BOSS CMASS data. Santi Serrano, Pau Tallada
and Davide Piscia for help in developing and maintaining the web portal.
We acknowledge support from the MareNostrum supercomputer (BSC-CNS, www.bsc.es), grants 
AECT-2008-1-0009, AECT-2008-2-0011, AECT-2008-3-0010,
and Port d'Informaci\'o Cient\'ifica (www.pic.es) where the simulations were ran and stored. 
We acknowledge the use of the Gadget-2 code (www.mpa-garching.mpg.de/gadget) to
implement the N-body and the
FoF code from the University of Washington to find halos
(www-hpcc.astro.washington.edu).
Funding for this project was partially provided 
by the European
Commission Marie Curie Initial Training Network CosmoComp
(PITN-GA-2009 238356),  the Spanish
Ministerio de Ciencia e Innovacion (MICINN), research
projects 200850I176,AYA-2009-13936, AYA-2012-39559, AYA-2012-39620,
Consolider-Ingenio CSD 2007-00060 and project SGR-1398
from Generalitat de Cata- lunya. MC acknowledges support from the Ramon
y Cajal MICINN program.
  
\bibliography{mnII.bib}




\end{document}